\documentclass[%
 reprint,
 amsmath,amssymb,
 aps,
prb,
]{revtex4-2}

\usepackage{graphicx}% Include figure files
\usepackage{dcolumn}% Align table columns on decimal point
\usepackage{bm}% bold math

\begin{document}

\title{On the dynamics of the Meissner and the Becker-London
  effects}

\author{Peter Marko\v{s} and Richard Hlubina}

\affiliation{Department of Experimental Physics, Comenius University,
  Mlynsk\'{a} Dolina F2, 842 48 Bratislava, Slovakia}

\date{\today}

\begin{abstract}
It is generally accepted that the most fundamental property of a
superconductor is that it exhibits the Meissner effect. Of similar
importance is the Becker-London effect, i.e. generation of magnetic
field inside a rotating superconductor. Hirsch has recently pointed
out that, within the conventional theory of superconductivity, the
question about how these effects are generated dynamically has not
even been asked yet. Here we fill in this gap in the literature by a
detailed study of the evolution of the electromagnetic field for both
of these effects. To this end, we solve the Maxwell equations
supplemented by the simplest conventional constitutive equation for a
superconductor, namely the London equation.  We demonstrate that,
contrary to the expectations of Hirsch, the conventional theory does
correctly describe the dynamics of both, the Meissner and the
Becker-London effect. We find that the dynamics of the studied
processes is quite rich and interesting even at this level of
description.

\end{abstract}
\maketitle

\section{Introduction}
This work is motivated by a series of papers by Hirsch, 
summarized in a recent semi-popular book \cite{Hirsch20}. Hirsch
argues that the conventional theory of superconductivity
\cite{Schrieffer64,deGennes66,Tinkham04,BCS11} is wrong or, at the
very least, incomplete.

Here we do not discuss the microscopic aspects of those papers.
Instead, we focus only on what we believe are two most fundamental
points of Hirsch's critique, namely on his discussion of the Meissner
effect \cite{Hirsch16annals,Hirsch20epl,Hirsch25} and of magnetic
field generation inside rotating superconductors
\cite{Hirsch14scripta,Hirsch19annalen}. The latter effect will be
called here, following the suggestion by Hirsch \cite{Hirsch20}, the
Becker-London effect. Hirsch argues that, in type-I superconductors,
the dynamics of none of these effects is addressed by the conventional
picture that a superconductor hosts a coherent condensate of the
Cooper pairs \cite{Schrieffer64,deGennes66,Tinkham04,BCS11}.

As regards the Meissner effect, Hirsch distinguishes between two
experimental protocols \cite{Hirsch16annals,Hirsch20epl,Hirsch25}.  He
first describes what we will call the ``cool-first protocol'': the
metal is initially cooled down to the superconducting state in a
vanishing field, and only afterwards a finite magnetic field is
switched on around it. In the cool-first protocol, Hirsch shows that
conventional theory is perfectly capable of describing the Meissner
effect: the growing magnetic field generates, by Faraday induction, an
electric field, this field accelerates the supercurrent, which finally
screens the $B$ field.

Within the second protocol considered by Hirsch, which we will call
the ``field-first protocol'', the metal is first placed in a finite
magnetic field and only afterwards it is cooled into the
superconducting state.  Experiments tell us that also in this case the
final state of the process is a superconductor which expels the
$B$-field \cite{Meissner33}.  In other words, the state of the
superconductor (in a sufficiently weak field) does not depend on its
history. This is of course the celebrated Meissner effect.

Hirsch admits that the standard theory does describe the {\it final
  state} of the field-first protocol correctly.  However, he argues
that the conventional theory does not describe the {\it process} which
leads from the initial to the final state.  Instead he finds that,
according to the conventional theory, the correct final state of the
field-first protocol should be a state with the same  $B$-field
as in the normal state.

More concretely, Hirsch argues that the conventional theory faces the
following four serious problems, when dealing with the field-first
protocol \cite{Hirsch16annals,Hirsch20epl,Hirsch20,Hirsch25}:

(i) the net force acting on the electrons at the moment when the
normal metal starts turning into a superconductor vanishes, because,
at this moment, no electric field is present in the sample. Therefore
the electrons should not start moving.

(ii) even if the electrons do start moving so that they screen the
external $B$-field, the Faraday induction will generate an $E$-field
which is oriented against the electronic current, thereby stopping the
electronic motion.

(iii) the superconducting body, if it is allowed to, rotates in an
opposite direction with respect to the direction dictated by the
Faraday $E$-field.

(iv) when a superconductor transforms to a normal metal in presence of
a finite magnetic field, the velocity of the Cooper pairs does not
decrease at the transition point. Rather, the supercurrent converts
into the normal current, which stops by a dissipative
process. Therefore the phase transformation process can not happen in
a dissipationless manner.

The second example where Hirsch identifies problems of the
conventional theory deals with the Becker-London effect, i.e. with
magnetic field generation inside rotating superconductors
\cite{Hirsch14scripta,Hirsch19annalen}.  Although this effect is not
as well known as the Meissner effect, it is well documented for a
broad spectrum of superconductors, see
\cite{Tate89,Verheijen90,Sanzari96} and references therein.

When discussing the Becker-London experiment, Hirsch again
distinguishes between two experimental protocols: in the cool-first
protocol, one first cools down the sample and only afterwards sets the
sample into rotation. In the rotate-first protocol, the order of
operations is reverted.

The cool-first protocol has been theoretically studied long ago
\cite{Becker33} with the conclusion that, when the superconductor is
treated as an ideal conductor, in its interior a finite field
develops, with a magnitude which depends only on the rotation
frequency and the electron mass and charge. London argues, by means of
an analogy to the Meissner effect, that the same field should be
generated also in the rotate-first protocol \cite{London61}.

Hirsch argues that, if they were treated within conventional theory,
the two protocols would lead to different outcomes of the
Becker-London experiment \cite{Hirsch14scripta,Hirsch19annalen}.
According to him, the cool-first protocol does produce the effect, but
within the rotate-first protocol no field should be generated, at
variance with experiments. The reason is essentially the same as in
the case of the Meissner effect: when a rotating normal metal enters
the superconducting state, there exists no driving force acting on the
superconducting electrons. Thus they should remain rotating along with
the ions as in the normal state, and no field should be generated.

To summarize, Hirsch argues that the conventional theory of
superconductivity violates Newton's law, the Maxwell electrodynamics,
the law of inertia, as well as the second law of thermodynamics
\cite{Hirsch20,Hirsch16annals,Hirsch20epl,Hirsch25,Hirsch14scripta,Hirsch19annalen}.
We are convinced that this criticism is serious enough to motivate a
thorough investigation of the {\it processes} leading from the initial
to the final states in both the Meissner and the Becker-London
experiments. We are not aware of a comprehensive study of this type.

Some exotic answers to the questions posed by Hirsch did appear
recently, however. For instance, Nikulov \cite{Nikulov22} suggests
that, instead of the conventional theory of superconductivity, it is
the second law of thermodynamics which is not universally
valid. Hirsch responds that ``the Meissner effect is consistent with
the second law of thermodynamics provided that a mechanism exists for
the supercurrent to start and stop without generation of Joule heat''
\cite{Hirsch25}, but he adds that the conventional theory of
superconductivity does not provide such a mechanism. Alternatively,
Koizumi suggests that the Meissner effect can be made reversible ``if
the theory is so modified that supercurrent generation is due to a
nontrivial Berry connection arising from many-body effects''
\cite{Koizumi20,Koizumi21}.

The approach of this paper to the questions posed by Hirsch is, in a
sense, exactly opposite. Rather than concentrating on exotic answers,
we will assume that the processes to be studied can be described by
well established tools: the Maxwell equations supplemented by the
simplest constitutive equation for the supercurrent of the
conventional theory of superconductivity, namely by the London
equation.  Throughout we assume that the superconductor is of type I
and we perform a complete numerical study of the temporal evolution of
the electromagnetic field for the two problems mentioned: the Meissner
effect and the Becker-London effect.  Both experimental protocols will
be studied in each case.

The outline of our paper is as follows. In
Section~II we start by presenting the basic equations of the theory.
In order to illuminate the essence of the studied phenomena, in the
rest of this paper the theory will be applied only to the simplest
geometries: a superconducting plate in a parallel $B$-field when
studying the Meissner effect in Section~III, and a superconducting
cylinder when describing the Becker-London effect in Section~IV.

We will show that the temporal evolution of the fields is in fact
quite different for the complementary protocols, as correctly
predicted by Hirsch. However, we will also show that the conventional
theory does correctly describe both the Meissner effect and the
Becker-London effect, irrespective of the chosen protocol.

%%%%%%%%%%%%%%%%%%%%%%%%%%%%%%%%%%%%%%%%%%%%%%%%%%%%%%%%%%%%%%%%%%%%%%%%%%%
\section{London electrodynamics}
The microscopic Maxwell equations governing the temporal evolution of
the electric and magnetic fields read
\begin{eqnarray}
\frac{1}{c^2}\frac{\partial{\bf E}}{\partial t}&=&
\nabla \times {\bf B}-\mu_0{\bf j},
\label{eq:Maxwell_E}
\\
\frac{\partial{\bf B}}{\partial t}&=&
-\nabla \times {\bf E}.
\label{eq:Maxwell_B}
\end{eqnarray}
Here ${\bf j}={\bf j}_{\rm ext}+{\bf j}_s$ is the total current
density, which is a sum of the externally applied currents ${\bf
  j}_{\rm ext}$ in the coils and the currents ${\bf j}_s$ flowing
inside the studied media.  The additional two Maxwell equations are
$\nabla\cdot{\bf E}=\rho$, where $\rho$ is the total charge density,
and $\nabla\cdot{\bf B}=0$.  The electromagnetic field will be
represented in the standard way in terms of the scalar potential
$\phi$ and of the vector potential ${\bf A}$ as ${\bf B}=\nabla \times
{\bf A}$ and ${\bf E}=-\nabla\phi-\partial{\bf A}/\partial t$.

In order to arrive at a closed set of equations, we need to specify
also the constitutive equations for the studied media, which determine
the currents ${\bf j}_s$ in terms of the electromagnetic field.
Throughout this manuscript we will neglect the small currents due to
normal electrons and concentrate on the supercurrents only. When
discussing thermodynamic aspects of the studied processes, the effects
caused by the normal electrons will be treated perturbatively at the
end of the calculations.

Therefore we assume that the current ${\bf j}_s$ is given by the
London equation \cite{London48,Ginzburg50}
\begin{equation}
  \mu_0{\bf j}_s=-\frac{f}{\lambda^2}
     \left({\bf A}+\frac{\hbar}{2e}\nabla\theta\right),
\label{eq:London}
\end{equation}
where $f({\bf x},t)$ is the superconducting fraction, $\lambda$ is the
penetration depth, $\theta({\bf x},t)$ is the phase of the condensate,
and $-e$ with $e>0$ is the electron charge
\cite{note_GL}. Non-superconducting media can be described by the same
formula by simply setting $f=0$.

The Maxwell equations coupled with Eq.~\eqref{eq:London} constitute
what we will call the London electrodynamics.

As is well known, the equations of the London electrodynamics are
invariant under the gauge transformation from the triplet of fields
$\phi,{\bf A},\theta$ to the triplet $\phi'=\phi-\partial\chi/\partial
t$, ${\bf A'}={\bf A}+\nabla\chi$, and
$\theta'=\theta-(2e/\hbar)\chi$, where $\chi({\bf x},t)$ is an
arbitrary field.

In this paper we will consider only situations in which charged
capacitors carrying external charge are absent. Moreover, in the geometries
to be studied we will assume that the electron charge density is
constant throughout the material and precisely opposite to the nuclear
charge density, so that the overall charge density $\rho=0$.
This is a standard assumption which differs from the phenomenology
proposed by Hirsch \cite{Hirsch03}.
Furthermore, we will assume that the superconducting bodies are
electrically isolated, and therefore we require that at their surface
${\bf j}_s\cdot{\bf n}=0$, where ${\bf n}$ is the surface normal.

Throughout this manuscript, we will work in the standard, so-called London
gauge: for the scalar potential we take $\phi=0$ and we assume that
$\nabla\cdot {\bf A}=0$. Furthermore, at the surface of the
superconductor, we require ${\bf A}\cdot{\bf n}=0$.  With this choice
of the potentials, following \cite{London48} one can show that for
simply connected superconductors the condensate phase $\theta$ has to
be constant, so that the last term in~\eqref{eq:London} does not
appear.  Moreover, from the requirement that the overall charge
density vanishes, $\rho=0$, it follows that the currents inside the
superconductor are purely transverse, $\nabla\cdot{\bf j}_s=0$.  In
the London gauge, this leads to the requirement that the vectors
$\nabla f$ and ${\bf A}$ are perpendicular. In the geometries to be
studied in this paper, this requirement is trivially satisfied.

Keeping in mind that in the general case the superconducting fraction
is a function of space and time, $f=f({\bf x},t)$, and taking the curl
of~\eqref{eq:London}, we obtain
\begin{equation}
\nabla\times {\bf j}_s=-\frac{1}{\mu_0\lambda^2}
\left[f{\bf B}+ \nabla f\times{\bf A}\right].
\label{eq:screening}
\end{equation}
Similarly, taking the time derivative of~\eqref{eq:London}, we find
\begin{equation}
\frac{\partial{\bf j}_s}{\partial t}=\frac{1}{\mu_0\lambda^2}
\left[f{\bf E}-\frac{\partial f}{\partial t}{\bf A}\right].
\label{eq:acceleration}
\end{equation}
In textbooks, one usually considers only the case when the
superconducting fraction $f$ is constant in space and time.  In that
case only the directly measurable fields ${\bf E}$ and ${\bf B}$
appear on the right-hand sides of
Eqs.~(\ref{eq:screening},\ref{eq:acceleration}).

However, when considering the {\it process} of transformation from the
normal to the superconducting state and vice versa, at the very least
the temporal evolution of $f$ has to be taken into account. In such
case the second term in the acceleration
equation~\eqref{eq:acceleration} does not vanish and, as we will see,
may become important.

Let us note that, taking the curl of Eq.~\eqref{eq:Maxwell_E} and
making use of~Eqs.(\ref{eq:Maxwell_B},\ref{eq:London}), the London
electrodynamics implies that the magnetic field satisfies the
following Proca-like equation 
\begin{equation}
\left[\frac{1}{c^2}\frac{\partial^2}{\partial t^2}+\frac{f}{\lambda^2}
-\bigtriangleup\right]{\bf B}=
\nabla\times \mu_0{\bf j}_{\rm ext}+\frac{1}{\lambda^2}\nabla f\times {\bf A}.
\label{eq:Klein_Gordon_general}
\end{equation}
One observes that, in the important special case when the
superconducting fraction $f$ is spatially uniform, only the external
currents flowing in the coils appear as a driving term on the
right-hand side of Eq.~\eqref{eq:Klein_Gordon_general}.

%%%%%%%%%%%%%%%%%%%%%%%%%%%%%%%%%%%%%%%%%%%%%%%%%%%%%%%%%%%%%%%%%%%%%%%%%%%
\section{Dynamics of the Meissner effect}
The goal of this Section is to consider in detail the temporal
evolution of the magnetic field within both experimental protocols for
the Meissner effect discussed by Hirsch. The calculations will be
performed numerically within the London electrodynamics introduced in
the previous Section.

\begin{figure}[]
\includegraphics[width = 7. cm]{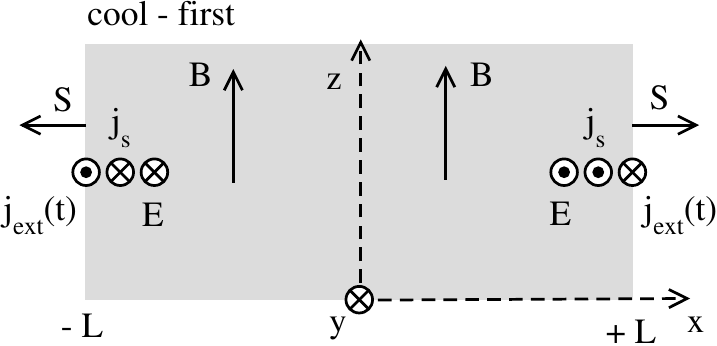}
\includegraphics[width = 7. cm]{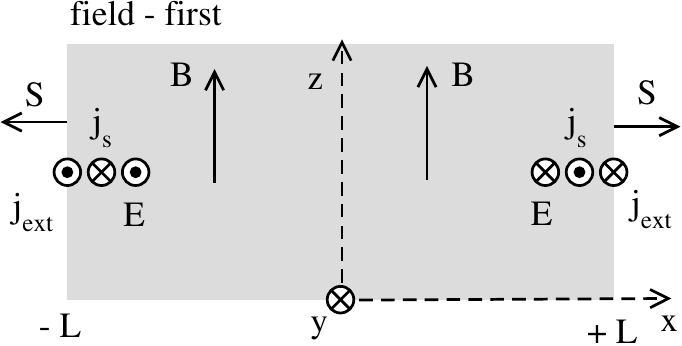}
\caption{Cut through the plate (shaded region) by the plane
  $yz$. Shown is the orientation of the relevant fields at an
  intermediate stage of the process leading to the Meissner effect, as
  obtained from the numerical solution. The crosses (dots) indicate
  vectors oriented into (out of) the paper plane.}
\label{fig:plate_scheme}
\end{figure}

In order to demonstrate the essence of the phenomena, we consider the
simplest possible geometry, namely that of a plate perpendicular to
the $x$ axis, see Fig.~\ref{fig:plate_scheme}.  The thickness of the
plate is $2L$ and we assume that $L\gg\lambda$.  The plate can be
either normal or superconducting. Outside the plate, i.e. for $|x|>L$,
there is vacuum and therefore $f=0$ in this region.

For definiteness, we assume that the magnetic field points along the
$z$ axis and that it depends only on the $x$ coordinate, ${\bf
  B}=(0,0,B(x))$. We also assume that the currents and the electric
field point in the $y$-direction and are given by ${\bf
  j}=(0,j(x),0)$, ${\bf E}=(0,E(x),0)$. Finally, in the London gauge
${\bf A}=(0,A(x),0)$.  This choice of geometry implies that the
equations of the London electrodynamics simplify to
\begin{eqnarray}
\frac{1}{c^2}\frac{\partial E}{\partial t}&=&
-\frac{\partial B}{\partial x}-\mu_0j_{\rm ext}+\frac{f(x,t)}{\lambda^2}A,
\label{eq:Maxwell_plate_E}
\\
\frac{\partial B}{\partial t}&=&-\frac{\partial E}{\partial x},
\label{eq:Maxwell_plate_B}
\\
\frac{\partial A}{\partial t}&=&-E,
\label{eq:Maxwell_plate_A}
\end{eqnarray}
where $j_{\rm ext}$ is the current in the ``coils'' around the plate,
which generate the magnetic field inside the plate. 

Note that, once the initial values of the fields $B(x,0)=\partial
A(x,0)/\partial x$ and $E(x,0)$ are known,
equations~(\ref{eq:Maxwell_plate_E},\ref{eq:Maxwell_plate_B},\ref{eq:Maxwell_plate_A})
govern the future temporal evolution of the fields $E$, $B$, and $A$.

In what follows we will consider the case when the external current is
spatially odd, $j_{\rm ext}(-x,t)=-j_{\rm ext}(x,t)$. It can be shown
that in this case also the fields $j(x,t)$, $E(x,t)$ and $A(x,t)$ are
spatially odd, while $B(x,t)$ is even.

\subsection{Switching on the B field in a superconductor}
Let us start with the cool-first protocol, i.e. let us assume that the
plate is always in the superconducting state and therefore $f=1$ in
the plate interior. We assume that the external current $j_{\rm ext}$
identically vanishes for all times $t<0$, and therefore the initial
electric and magnetic fields at time $t=0$ vanish as well, $E(x,0)=0$
and $B(x,0)=0$.

At time $t=0$, we start switching on the external current and we
assume that for $t>0$
\begin{equation}
\mu_0j_{\rm ext}(x,t)=B_0\left[\delta(x-L)-\delta(x+L)\right] g(t),
\end{equation}
where the function $g(t)$ describes the switch-on process, i.e. we
require that it grows monotonically from $g(0)=0$ to $g(\infty)=1$.
As a simple example, in what follows we take $g(t)=\tanh(t/\tau)$,
where $\tau$ measures the duration of the switch-on process. We shall
assume that $\tau\gg L/c$, so that this process is slow when compared
with the time needed by the electromagnetic wave to traverse the
plate.

In the limit of sufficiently long times $t\gg \tau$ the magnetic field
in the plate interior should exhibit the Meissner effect, and
therefore $B(x)$ should be given by
\begin{equation}
B(x)=B_0\frac{\cosh(x/\lambda)}{\cosh(L/\lambda)}.
\label{eq:meissner_plate}
\end{equation}
Since $L\gg\lambda$, the screening of the magnetic field is well
pronounced.  Our goal here is to study the temporal evolution of the
fields during the switch-on process.

\begin{figure}[]
\includegraphics[width = 7. cm]{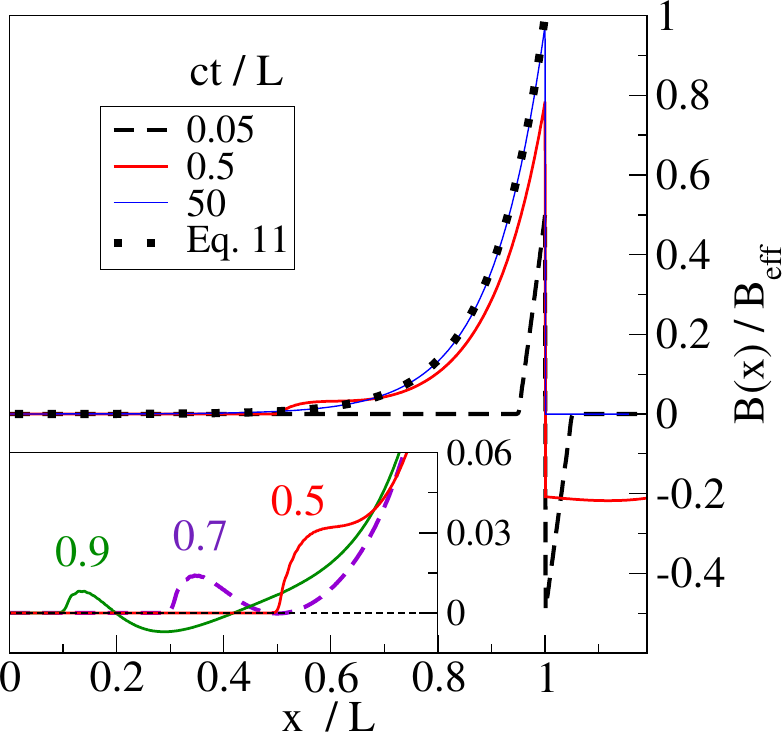}
\caption{Temporal evolution of the magnetic field profile in the plate
  (for $\lambda/L=0.1$ and $c\tau/L=20$) for the cool-first
  protocol. The dots correspond
  to~\eqref{eq:meissner_plate} with $B_{\rm eff}=B_0$.
  The inset shows the field profiles for times
  $\lambda/c<t<L/c$, when the field in the middle of the plate is
  still identically zero. }
\label{fig:sn_B}
\end{figure}

As shown in Fig.~\ref{fig:sn_B}, several temporal regimes can be
distinguished in the numerical solution. At the shortest time scales
$t<\lambda/c$, the field penetrates the plate as though the plate was
not superconducting. For instance in the vicinity of the right edge of
the plate $x\lesssim L$, we find that $B(x,t)\approx B_0t_{\rm
  ret}/(2\tau)$, $E(x,t)=-cB(x,t)$, and $A(x,t)\approx B_0c t_{\rm
  ret}^2/(4\tau)$, where the retarted time is given by $t_{\rm
  ret}=t-(L-x)/c$. At distances larger than $ct$ from both ends of the
plate, all fields are identically zero, as required by the finite
value of $c$.

The most interesting regime is $\lambda/c<t<L/c$. In this regime, the
interior region where all fields are identically zero is still
present, but close to the plate surface the superconductor starts to
screen the magnetic field. The screening current is driven by the
non-vanishing value of the vector potential, which happens to be
positive in the vicinity of $x=L$, leading to a negative value of the
screening current, as is necessary for the Meissner effect, see
Fig.~\ref{fig:plate_scheme}.  In agreement with the analysis of
Hirsch, also the electric field is negative close to the right
surface, and, according to~\eqref{eq:acceleration}, the absolute value
of the screening current therefore grows.

At times $L/c<t<\tau$, the wave fronts from both sides of the plate
merge and all fields penetrate the whole interior of the plate. We
find that the magnetic field profile quickly approaches that predicted
by~\eqref{eq:meissner_plate} with $B_0$ replaced by a time-dependent
effective field
\begin{equation}
B_{\rm eff}(t)=B_0 g(t)=B_0 \tanh(t/\tau).
\label{eq:B_eff}
\end{equation}
Note that the resulting field $B(x,t)$ solves the Proca-like
equation~\eqref{eq:Klein_Gordon_general} in which the time-derivative
term on the left-hand side, caused by the Maxwell displacement and
proportional to $c^{-2}$, is neglected. Finally, for very long times
$t\gg\tau$ the magnetic field approaches the expected
form~\eqref{eq:meissner_plate}.

{\it Conservation of energy}. As is well known, from the Maxwell
equations one can derive a continuity equation for the total energy of
the system charges + field \cite{Jackson99}. In our case this equation
reads
\begin{equation}
\frac{\partial u}{\partial t}+\frac{\partial S}{\partial x}+jE=0,
\label{eq:energy_conservation_general}
\end{equation}
where $u=(E^2/c^2+B^2)/(2\mu_0)$ is the field energy density and
$S=EB/\mu_0$ is the $x$-component of the Poynting vector ${\bf
  S}=(S,0,0)$ describing the flow of the field energy. Taking the
integral $\int_0^\infty dt\int_0^{L+\delta}dx$ (where $\delta=0^+$) of
this equation, one can show that conservation of energy implies that
\begin{equation}
W=\Delta U+\int_0^\infty dt S(L+\delta,t).
\label{eq:energy_conservation}
\end{equation}
Here $W$ is the work (per unit area) supplied by the battery to the
right coil,
\begin{equation}
W=-\frac{B_0}{\mu_0}\int_0^\infty dt \tanh(t/\tau) E(L,t),
\label{eq:W_plate}
\end{equation}
and $\Delta U=U(\infty)-U(0)$ is the change of the energy $U(t)$ of
the right half-plate (per unit area) during the switch-on process,
where 
\begin{equation}
  U=\frac{1}{2\mu_0}\int_0^L dx
  \left[\frac{1}{c^2}E^2+B^2+\frac{f}{\lambda^2}A^2\right].
\end{equation}
Note that $U$ is the sum of the field energy and of the kinetic energy
of the currents described by the last term. Evaluating $U(\infty)=U_s$
by taking into account that at the end of the process the electric
field $E=0$ and the magnetic field is described
by~\eqref{eq:meissner_plate}, one finds that
\begin{equation}
U_s=\frac{B_0^2\lambda}{2\mu_0}\tanh(L/\lambda). 
\label{eq:energy_us}
\end{equation}
Taking furthermore into account that the initial energy $U(0)=0$, we
obtain $\Delta U=U_s$.

The result~\eqref{eq:energy_conservation} has a simple interpretation:
part of the work $W$ supplied by the battery is used to increase the
energy of the plate, while
\begin{equation}
W_{\rm rad}=\int_0^\infty dt S(L+\delta,t),
\quad
S(L+\delta,t)=E^2(L,t)/(\mu_0 c)
\label{eq:W_rad_plate}
\end{equation}
measures the energy (per unit area) radiated towards $x=+\infty$ from
the plate as an electromagnetic pulse.

\begin{figure}[t]
\includegraphics[width = 7. cm]{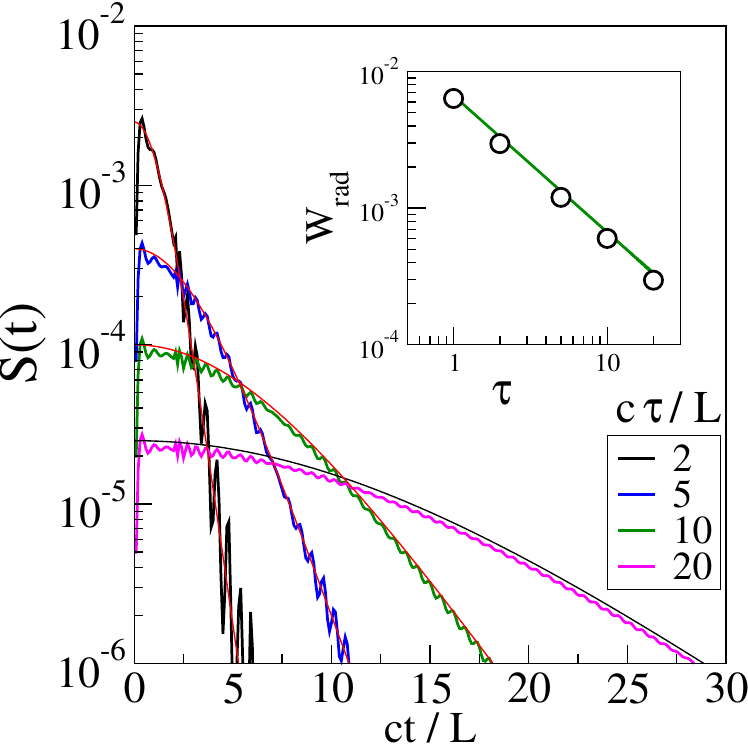}
\caption{Temporal evolution of the Poynting vector in close vicinity
  of the plate (for $\lambda/L=0.1$ and several values of $\tau$) for
  the cool-first protocol. The lines are analytical estimates in the
  large-$\tau$ limit, described in the text. The Poynting vector is
  proportional to $\cosh^{-4}(t/\tau)$ and is measured in units of
  $cB_0^2/\mu_0$. The inset shows the scaling of the total radiated
  energy $W_{\rm rad}$ (in units of $B_0^2L/\mu_0$) with $\tau$. }
\label{fig:sn_poynting}
\end{figure}

The radiative losses can be minimized by increasing the time $\tau$,
as can be expected on general grounds and as confirmed in
Fig.~\ref{fig:sn_poynting} by an explicit numerical calculation.  In
the limit $\tau\rightarrow \infty$, conservation of energy therefore
implies that the work supplied by the battery is $W=U_s$.

This result can be checked by an explicit calculation as follows. Let
us assume that, for large values of $\tau$, $B(x,t)$ is given for all
times by~\eqref{eq:meissner_plate} with $B_0$ replaced by $B_{\rm
  eff}(t)$.  From here we can calculate the vector potential
$A(x,t)=\int_0^x dx' B(x',t)$, as well as the electric field
$E(x,t)=-\partial A(x,t)/\partial t$.  Taking the integral in
Eq.~\eqref{eq:W_plate}, we find that $W=U_s$, as expected. If we
furthermore assume that $L\gg \lambda$, by a similar calculation we
also find that the total radiated energy is $W_{\rm rad}\approx
2B_0^2\lambda^2/(3\mu_0 c\tau)$.  As shown in the inset of
Fig.~\ref{fig:sn_poynting}, this result is in reasonable agreement
with numerics down to very small values of $\tau$.

So far we have neglected the presence of normal electrons.  But, at a
finite temperature, they are necessarily present.  If we do include
them as a small perturbation, we can use the already calculated
electric field $E(x,t)$ in order to estimate the Joule heat developed
in the transition process. To this end, we have to evaluate the
expression
\begin{equation}
Q_J=\int_0^\infty dt P_J(t),
\quad
P_J(t)=\sigma\int_0^L dx E^2(x,t),
\label{eq:joule_heat}
\end{equation}
where $\sigma$ is the conductivity of the normal electrons.  We find
that $Q_J\approx \sigma B_0^2 \lambda^3/(3\tau)$.  This means that the
radiated energy $W_{\rm rad}$ and the Joule heat $Q_J$ have two
important properties in common: both scale with $1/\tau$ and neither
of them depends on $L$, i.e. none of them is extensive. This means
that both of them can be safely ignored in thermodynamic
considerations.

{\it Conservation of momentum}. Let us denote the momentum density of
the ions in the $y$ direction as $\pi_{\rm nucl}$. As shown in
Appendix~A, assuming that the normal electrons are tightly bound to
the nuclei, the acceleration of the ions is given by
\begin{equation}
\frac{\partial \pi_{\rm nucl}}{\partial t}
=nf eE-\frac{\partial (nf)}{\partial t}p_s,
\label{eq:acceleration_ions}
\end{equation}
where $p_s=eA$ is one half of the Cooper pair momentum.  The first
term in~\eqref{eq:acceleration_ions} is the Lorentz force acting on
the system 'nuclei+normal electrons' with total charge density $nfe$,
while the second term (to be discussed in more detail later) describes
momentum exchange between the ions and the condensate. In the present
case of the cool-first protocol, only the first term is present.
Therefore, if they were allowed to, the ions would accelerate in the
same direction as the electric field is oriented.

A word of caution is in place here: Eq.~\eqref{eq:acceleration_ions}
can not hold locally, since the ions form a rigid body. Rather, this
equality can only have consequences for the motion of the body as a
whole. This effect is not considered here, however, and we assume that
the plate is firmly positioned in space.

\subsection{Uniform condensate formation in a finite B field}
In order to describe the field-first protocol, we take the currents in
the coils as time-independent,
\begin{equation}
\mu_0j_{\rm ext}(x)=B_0\left[\delta(x-L)-\delta(x+L)\right].
\label{eq:current_ns}
\end{equation}
We assume that at negative times $t<0$ the plate is normal, and
therefore $f=0$ for all $x$. This implies that the initial magnetic
field at time $t=0$ is $B(x,0)=B_0=\mu_0 H$ inside the plate, while
$B(x,0)=0$ outside. Moreover, the initial electric field $E(x,0)=0$
for all $x$.

The question we ask is how does the electromagnetic field develop as a
function of time, if for positive times $t>0$ the superconducting
fraction $f(x,t)$ inside the plate is switched on. To start with, for
pedagogical reasons, we assume that $f(t)$ does not depend on the
position $x$ and is given by $f(t)=\tanh(t/\tau)$. We shall again
assume that the switch-on process is slow, $\tau\gg L/c$. Note that we
assume that the ``superconducting fraction'' at the end of the process
satisfies $f(\infty)=1$. This does not mean that the superconductor is
at zero temperature, however. The reason is that for the value of
$\lambda$ in~\eqref{eq:London} we take the (finite) penetration depth
at the superconducting side of the phase transition at
$T_c(H)$. Therefore $f$ should rather be understood as the ``degree to
which the superconducting order has been established''.

\begin{figure}[t]
\includegraphics[width = 7. cm]{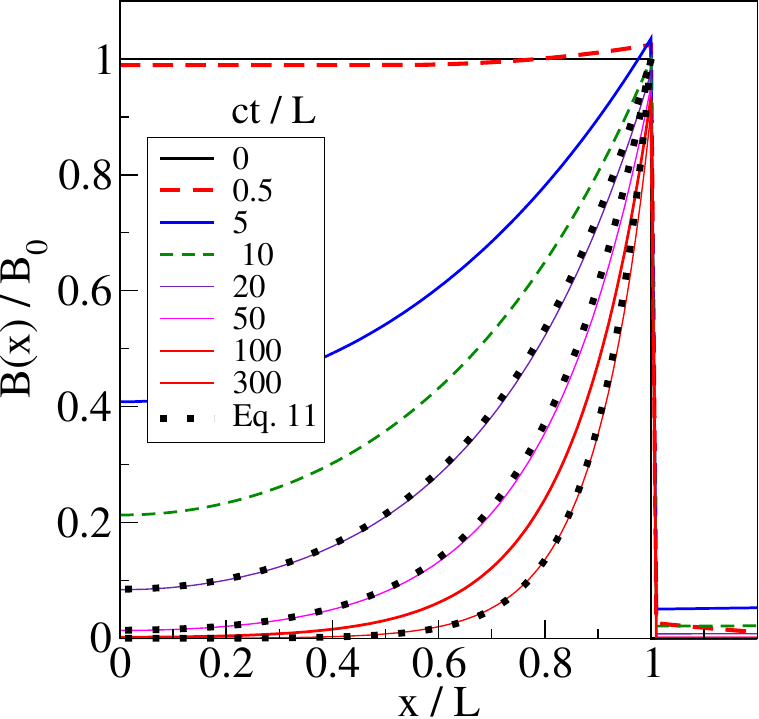}
\caption{Temporal evolution of the magnetic field profile in the
  sample (for $\lambda/L=0.1$ and $c\tau/L=200$) for the field-first
  protocol.  The dots correspond to~\eqref{eq:meissner_plate} with
  $\lambda$ replaced by $\lambda_{\rm eff}(t)$,
  see~\eqref{eq:lambda_eff}.}
\label{fig:ns_B}
\end{figure}

Figure~\ref{fig:ns_B} shows the magnetic field profiles obtained by a
numerical solution of the equations of the London electrodynamics for
such an experimental protocol.

\begin{figure}[t]
\includegraphics[width = 7. cm]{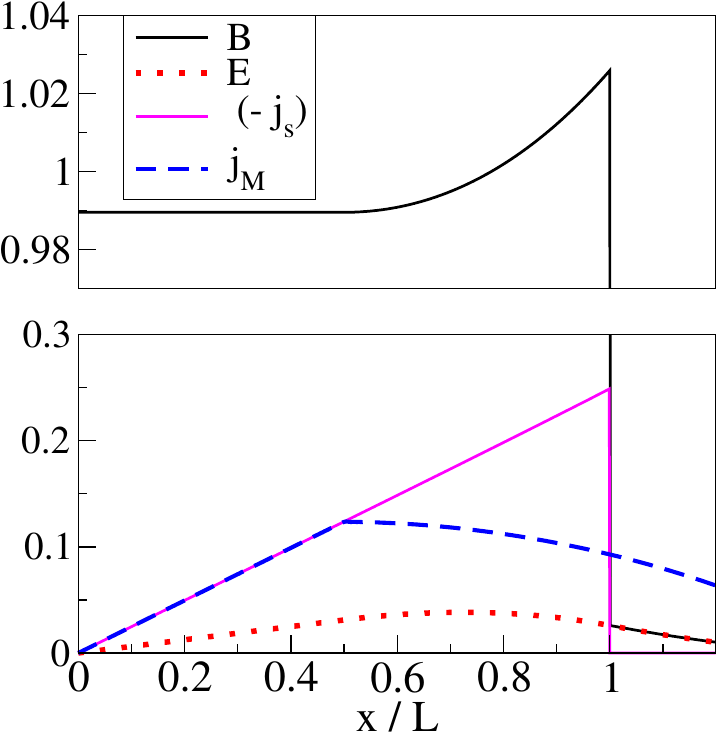}
\caption{Spatial profile of the relevant fields in the sample (for
  $\lambda/L=0.1$, $c\tau/L=200$ and at very short time $ct/L=0.5$) for
  the field-first protocol. The Maxwell displacement current is
  defined as $\mu_0 j_{\rm M}=c^{-2}\partial E/\partial t$. The
  magnetic field, electric field, and the current densities are
  measured in units of $B_0$, $cB_0$, and $B_0/(\mu_0 L)$,
  respectively.}
\label{fig:ns_polia}
\end{figure}

For the shortest times $t<L/c$, the vector potential is essentially equal to
$A\approx B_0x$, and, as a result, according to~\eqref{eq:London} a
finite supercurrent flows in the plate which screens the external
field, as shown in Fig.~\ref{fig:ns_polia}.  This solves Hirsch's 
problem (i) from the Introduction.

The supercurrent flow causes the magnetic field to decrease, as we
move from the sample surface towards the center of the plate.
However, we find that this decrease is limited only to distance $ct$
from the plate surface, and deeper inside the plate the magnetic field
is exactly constant.  As follows from~\eqref{eq:Maxwell_plate_E}, this
means that it is the Maxwell displacement current $\mu_0 j_{\rm
  M}=c^{-2}\partial E/\partial t$ which precisely compensates the
supercurrent deep in the plate interior, in perfect agreement with
Fig.~\ref{fig:ns_polia}.  The electric field is thus given here by
$E(x,t)\approx B_0xc^2t^2/(2\lambda^2\tau)$.  Furthermore, making use
of~\eqref{eq:Maxwell_plate_B}, we find that the magnetic field in the
plateau region is given by $B(x,t)\approx
B_0-B_0c^2t^3/(6\lambda^2\tau)$, showing the first hints of the
Meissner effect.

Before proceeding, let us note that although the electric field is
positive in the right half of the plate (in agreement with the Faraday
law), the supercurrent flows here in the negative direction, see also
Fig.~\ref{fig:plate_scheme}. Moreover, the absolute value of $j_s$
increases as a function of time. The explanation of this seemingly
paradoxical behavior is provided by the acceleration
equation~\eqref{eq:acceleration}: the usually absent second term in
this equation dominates. Therefore, since the superconducting fraction
$f$ grows as a function of time, the supercurrent accelerates in the
negative $y$ direction, as required by the Meissner effect. This
solves Hirsch's problem (ii) from the Introduction.

Moreover, by virtue of~\eqref{eq:acceleration_ions}, also the ions
would move in the negative $y$ direction, i.e. against the electric
field, if they were allowed to. This solves Hirsch's problem (iii). As
regards the microscopic mechanism of the momentum transfer between the
superconducting electrons and the ions, we nevertheless agree with
Hirsch: the condensate forms by annihilating pairs of normal electrons
with a finite net momentum $2p_s$ along $+y$. Thus there is a surplus
of normal electrons in the $-y$ direction. The normal electrons relay
the resulting surplus momentum to the lattice by scattering on
impurities and/or phonons.

Surprisingly, still for times $t<L/c$, the magnetic field right at the
plate surface grows as a function of time, generating outside the
plate an electromagnetic pulse caused by the changing field inside the
plate. In retrospect, it is easy to see that this must be so: the
electric field at the plate surface grows as a function of time, but
since it is continuous across $x=L$ and since the electric and
magnetic fields are related by $B=E/c$ outside the plate (see also
Fig.~\ref{fig:ns_polia}), the $B$-field on both sides of $x=L$, where
it is discontinuous, must grow with time as well.

We find (not shown) that the electromagnetic pulse peaks at times
$t\sim L/c$. At times $t> L/c$, the plateau of the magnetic field deep
inside the plate disappears and the field distribution approaches
the Meissner form~\eqref{eq:meissner_plate} with $\lambda$ replaced by
the effective penetration depth at time $t$ defined by
\begin{equation}
\lambda_{\rm eff}^{-2}(t)=\lambda^{-2}f(t)=\lambda^{-2}\tanh(t/\tau),
\label{eq:lambda_eff}
\end{equation} 
see Fig.~\ref{fig:ns_B}. Note that, exactly as in III.A, the resulting
field $B(x,t)$ solves the Proca-like
equation~\eqref{eq:Klein_Gordon_general} with neglected Maxwell
displacement current. Concomitantly, at times $t> L/c$ the
electromagnetic pulse decays.

A note is in place here. Since the magnetic field close to the sample
surface is larger than the critical field $B_0$, the superconducting
state is at most metastable in this spatial region. Here we assume
that, nevertheless, the condensate formation is not affected by this
fact, in agreement with experiments in clean samples. On the other
hand, it is well known that in dirty samples the flux expulsion may be
only partial even in type-I superconductors, see
e.g. \cite{Hendricks71,Kozhevnikov21}.  The treatment of such case is
beyond the scope of this paper.

{\it Conservation of energy.}  Starting from
\eqref{eq:energy_conservation_general} and taking the integral
$\int_0^\infty dt\int_0^{L+\delta}dx$ as in the previous subsection,
after some algebra we obtain \cite{note_conserv_en}
\begin{equation}
U_s+W_{\rm rad}=U_n+W_B-W.
\label{eq:energy_ns}
\end{equation}
The left-hand side of~\eqref{eq:energy_ns} is the total energy of the
final state, which includes also the energy of the electromagnetic
pulse. The right-hand side is the energy of the initial normal state
$U_n=B_0^2L/(2\mu_0)$, corrected for the work $W=B_0\int_0^\infty dt 
E(L,t)/\mu_0=2(U_n-U_s)$ done on the coils and for absorbed energy
$W_B$ needed to expel the magnetic field,
\begin{equation}
  W_B=\int_0^\infty dt P_B(t),
\quad
P_B(t)=\frac{1}{2\mu_0\lambda^2}
\int_0^L dx \frac{\partial f}{\partial t} A^2(x,t).
\label{eq:W_B}
\end{equation}
Note that $W_B$ can also be interpreted as the energy needed to join
the moving condensate.  The source of $W_B$ will be identified
shortly.

Similarly as in the previous subsection, the total radiated energy
$W_{\rm rad}$ scales as $1/\tau$ (not shown) and therefore, in the
limit of large switch-on times $\tau$, Eq.~\eqref{eq:energy_ns}
implies that $W_B=U_n-U_s$ \cite{note_W_B}.  Moreover, in the limit of
very thick plates, one can neglect $U_s$ with respect to $U_n$ and
therefore we find that $W_B=B_0^2L/(2\mu_0)$. Note that in this
limit~\eqref{eq:energy_ns} implies that $W=U_n+W_B$, which is in
perfect agreement with~\cite{Hirsch25}: one half of the work $W=2U_n$
done on the coils is supplied by the normal-state energy $U_n$, and
the other half by $W_B=U_n$.

Next let us compare our result~\eqref{eq:energy_ns} with the standard
thermodynamic considerations \cite{Tinkham04}.  To this end, let us
define the Gibbs free energy (per unit area of the plate) as
$G(T,H)={\cal E}-TS-\mu_0 HML$, where $H=B_0/\mu_0$ is the external
field, ${\cal E}$ is the internal energy, $S$ the entropy, and $M$ the
magnetization of the plate.  At the critical point the Gibbs free
energies of the normal and superconducting states are equal,
$G_s=G_n$, from where it follows that
\begin{equation}
{\cal E}_s={\cal E}_n-T(S_n-S_s)-\mu_0 H L(M_n-M_s).
\label{eq:thermodynamics_ns}
\end{equation}
This means that the transition from the normal to the superconducting
state is associated with emission of latent heat $Q=T \Delta
S=T(S_n-S_s)$ to the reservoir, and with supplying work $W=\mu_0
HL(M_n-M_s)$ to the coils. Since in the superconducting state
$M_s=-H$, and (neglecting the weak normal-state magnetism) $M_n=0$, we
find that $W=\mu_0H^2L$, in agreement with the electrodynamic
calculation.

In order to find the source of energy $W_B$, let us note that the full
energies ${\cal E}_s$, ${\cal E}_n$ of the superconducting and normal
states differ from $U_s$, $U_n$. In fact, we can set ${\cal
  E}_s=U_s-\Delta {\cal E}$ and ${\cal E}_n=U_n$, where $\Delta {\cal
  E}>0$ describes the lowering of energy in the superconducting state
due to condensate formation (in zero field). Comparing
Eqs.~(\ref{eq:energy_ns},\ref{eq:thermodynamics_ns}) we thus find that
$W_B=\Delta {\cal E}-T\Delta S=\Delta F$, where $\Delta F$ is the
difference of the Helmholtz free energies between the normal and
superconducting states in absence of the magnetic field.

This means that $W_B$ comes from the energy lowering $\Delta {\cal E}$
due to condensate formation, diminished by the latent heat $Q=T\Delta
S$ emitted to the reservoir in the transition process. Let us also
note in passing that the critical temperature $T_c(H)$ in the field
$H$ can be determined from $\Delta F(T_c)=\mu_0 H^2L/2$, as is well
known \cite{Tinkham04}.

Summarizing, we have demonstrated that the London electrodynamics does
correctly describe the magnetic field expulsion, if we postulate that
the superconducting fraction grows homogeneously. However, such a
scenario has the following serious problem. If we estimate the
magnetic field $B(x,t)$ by~\eqref{eq:meissner_plate} with $\lambda$
replaced by the effective penetration depth~\eqref{eq:lambda_eff},
similarly as in III.A we can easily predict also the fields $A(x,t)$
and $E(x,t)$. Making use of these estimates in~\eqref{eq:W_rad_plate}
and~\eqref{eq:joule_heat}, we find $W_{\rm rad}\approx 0.095 B_0^2
L^4/(\mu_0 c \tau \lambda^2)$ and $Q_J\approx \sigma B_0^2 L^5/(24
\tau \lambda^2)$.  These results are obviously unphysical, since both
quantities are more than extensive. In fact, also Hirsch has argued that
the homogeneous development of the condensate is energetically 
impossible.  In the next subsection we will
show how to cure this problem.

\subsection{Nucleation of condensate in a finite B field}
It is crucial to realize that the phase transition from the normal
metal to a type-I superconductor in a finite magnetic field is of
first order. From here it follows that the transition should proceed
via the standard nucleation mechanism. In other words, the
superconducting condensate should initially form in a small nucleus
and from there it should spread to the whole plate. In what follows we
therefore study the time evolution of the magnetic field, taking into
account both temporal and spatial dependence of the condensate
fraction $f(x,t)$.

\begin{figure}[t]
\includegraphics[width = 7. cm]{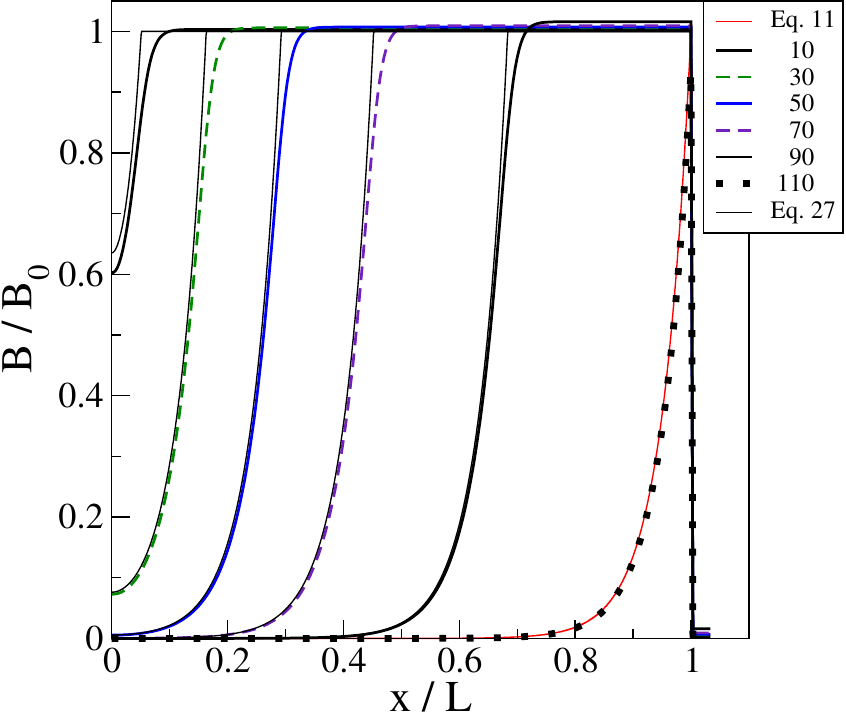}
\caption{Temporal evolution of the magnetic field profile in the plate
  in case when the superconducting phase with $\lambda/L=0.05$
  nucleates inside a normal metal.  The nucleation front at $h(t)$
  (defined in the text) moves outward from the plate center with
  diffusion constant $D/(cL) = 0.01$ for $h<h_0$ and with velocity
  $v_{\rm max}$ for $h>h_0$, where $h_0 = 0.23L$.  For the condensate
  fraction we take $f(x)=n(|x|-h)$, where $n(x)$ is the Fermi function
  with ``temperature'' $\delta=0.01L$.  Times are measured in units of
  $L/c$.}
\label{fig:plate_nucleus_ns}
\end{figure}

For the sake of simplicity, let us assume that superconductivity
nucleates at time $t=0$ in the center of the plate and at time $t$ the
nucleation fronts are at $\pm h(t)$. Thus the region $|x|<h(t)$ is
superconducting and the ``superfluid fraction'' is given by
$f(x,t)=\theta(h-|x|)$, where $\theta(x)$ is the Heaviside
function. In actual numerical calculations, instead of the Heaviside
function we take the Fermi function with a small smearing of the step.
This smearing should be at least comparable to the size of the Cooper
pair.

Let us furthermore assume, following \cite{Livingston69}, 
that the differential equation for the velocity
of the front is
\begin{equation}
v=\dot{h}=\frac{D}{2(L-h)}.
\label{eq:front_velocity}
\end{equation}
At the end of the calculation we will present an alternative
derivation of this equation. The solution to Eq.~\eqref{eq:front_velocity}
which satisfies the initial condition $h(0)=0$ reads
$h(t)=L-\sqrt{L^2-Dt}$. From here it follows that the velocity of the
front is $v(t)=D/(2\sqrt{L^2-Dt})$ and the duration of the nucleation
process is $t_{\rm max}=L^2/D$. Note that, at the end of the process,
the velocity of the front diverges. This is clearly unphysical and therefore
we limit the front velocity by $v_{\rm max}$. We assume that this
velocity is reached when the front is located at $L-h_0$, and thus
$h_0=D/(2v_{\rm max})$. At later times we assume that the front velocity does 
not increase any more, $v(t)=v_{\rm
  max}$.

For this choice of the function $f(x,t)$, we have solved the equations
of the London electrodynamics with the initial conditions $B(x,0)=B_0$
and $E(x,0)=0$ numerically. In order to nucleate superconductivity in
an originally normal metal, for the external field we have to take
$B_0<B_c$, so that the truly thermodynamically stable state is a
superconductor.

As in the simpler case of homogeneous
condensate formation, also in the present case we find that the
superconductor does exhibit the Meissner effect, as shown explicitly
in Fig.~\ref{fig:plate_nucleus_ns}.  

For future reference it is useful to find analytic approximations to
the obtained numerical results. If the diffusion constant $D$ is
sufficiently small, we can neglect the time derivative term
in~\eqref{eq:Klein_Gordon_general} and for the magnetic field
distribution at time $t$ we obtain the following estimate:
\begin{equation}
B(x,t)=B_0 \times {\rm min}
\left[\frac{\cosh(x/\lambda)}{\cosh(h/\lambda)},1\right].
\label{eq:plate_nucleus_B}
\end{equation}
As shown in Fig.~\ref{fig:plate_nucleus_ns}, this formula
describes the magnetic field reasonably well.
From~\eqref{eq:plate_nucleus_B} we can easily calculate the vector
potential $A(x,t)=\int_0^x dx' B(x',t)$ and also the electric field
$E(x,t)=-\partial A(x,t)/\partial t$. We find
\begin{equation}
E(x,t)={\rm sgn}(x) B_0v\tanh(h/\lambda)
\frac{\sinh({\rm min}(|x|,h)/\lambda)}{\cosh(h/\lambda)}.
\label{eq:plate_nucleus_E}
\end{equation}

\begin{figure}[t]
\includegraphics[width = 6. cm]{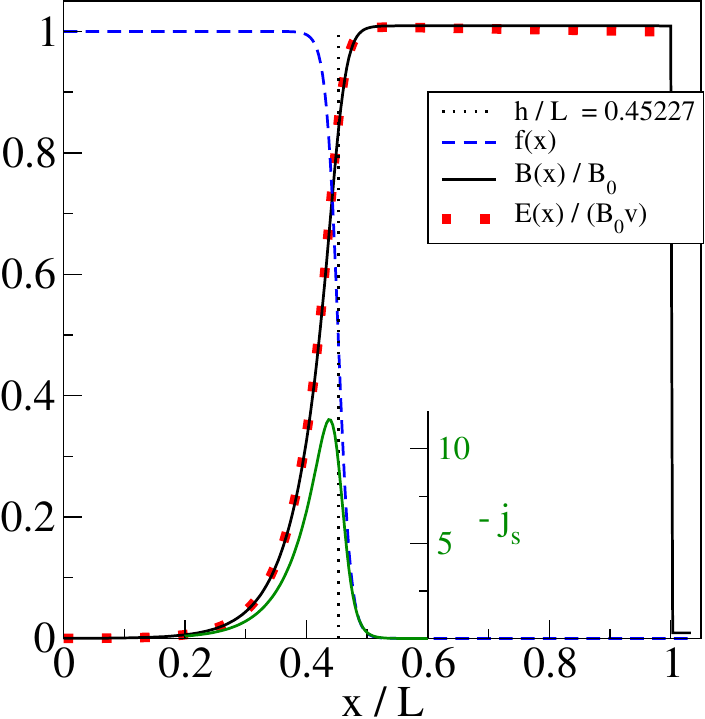}
\caption{Spatial profile of the magnetic field $B$, electric field
  $E$, condensate fraction $f$, and supercurrent $j_s$ (in units of
  $B_0/(\mu_0 L)$) in case when the nucleation front lies at $h\approx
  0.45 L$, for the same parameters as in Fig.~\ref
  {fig:plate_nucleus_ns}.}
\label{fig:plate_nucleus_polia}
\end{figure}

A qualitative explanation of why the magnetic field is expelled from
the right half of the plate is as follows.  Let us suppose that the
nucleation front is at $h(t)$.  It is true that the first term of the
acceleration equation~\eqref{eq:acceleration} predicts that in the
already superconducting part $0<x<h$ of the plate the absolute value
of the (negative) supercurrent decreases, since $\partial j_s/\partial
t>0$.  However, much more important is the second term
in~\eqref{eq:acceleration}, which generates a supercurrent with the
correct (negative) direction right at the front position, since
$\partial f/\partial t>0$ and thus $\partial j_s/\partial
t<0$. Therefore, as time proceeds, supercurrent which screens the
magnetic field is always pinned to the nucleation front, see
Fig.~\ref{fig:plate_nucleus_polia}.

The energy $W_{\rm rad}$ radiated in the transition
process is given by~\eqref{eq:W_rad_plate}. With the above estimates
of the fields we find that the Poynting vector at the plate surface is
$S(t)=B_0^2 v^2\tanh^4(h/\lambda)/(\mu_0 c)$.  From here it follows
that the total radiated energy is
\begin{equation}
W_{\rm rad}=\frac{B_0^2}{\mu_0 c}\int_0^L dh v \tanh^4(h/\lambda)
\approx \frac{B_0^2 D}{2\mu_0 c}\ln\left(\frac{{\rm e}L}{h_0}\right),
\label{eq:nucl_rad}
\end{equation}
where ${\rm e}$ is Euler's number.  In this estimate, we have
neglected the small contribution of $h\lesssim\lambda$ and we have set
$\tanh(h/\lambda)=1$.  The crucial point to observe here is that the
radiated energy is not extensive.

Let us also find an estimate of the Joule heat $Q_J$ generated in the
transition process by evaluating~\eqref{eq:joule_heat} with electric
fields taken from a calculation not involving $\sigma$. When estimating
the Joule power $P_J(t)$, one has to take into account that the
normal-electron conductivity is different in the normal metal and in
the superconductor: $P_J(t)=\int_0^L dx \sigma(x)E^2(x,t)$. In the
normal metal we take for $\sigma(x)$ the normal-state conductivity
$\sigma$, while in the superconductor $\sigma(x)$ is given by the
conductivity $\sigma'$ due to the normal fluid in the superconducting
state.  In what follows we will neglect the contribution proportional
to $\sigma'$, since its contribution to $Q_J$ turns out to be less
than extensive. 

\begin{figure}[t]
\includegraphics[width = 6. cm]{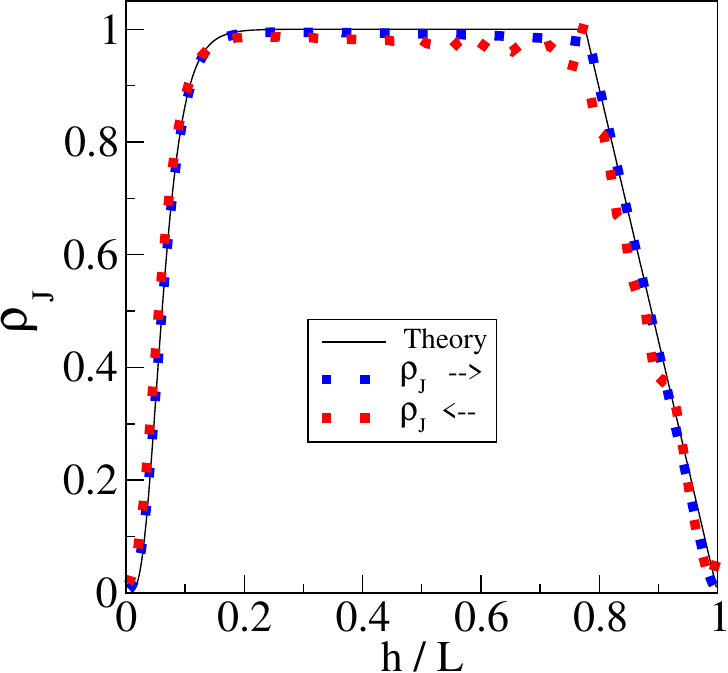}
\caption{The Joule loss density $\rho_J(h)$ in units of $\sigma
  B_0^2D/2$ for the nucleation processes discussed in III.C and III.D.
  The nucleation front at $h(t)$ (defined in the text) moves outward
  (arrow to the right) or inward (arrow to the left) with diffusion
  constant $D/(cL) = 0.01$ for $h<h_0$ and with velocity $v_{\rm max}$
  for $h>h_0$, where $h_0 = 0.23L$.  For the condensate fraction we
  take $f(x)=n(|x|-h)$, where $n(x)$ is the Fermi function with
  ``temperature'' $\delta=0.01 L$. The penetration depth in the
  superconducting phase is $\lambda/L=0.05$.}
\label{fig:plate_joule}
\end{figure}

Introducing the Joule loss density $\rho_J(h)$ via
$P_J(t)=v \rho_J(h)$, we find that $\rho_J(h)=\sigma B_0^2 v(L-h)
\tanh^4(h/\lambda)$. This estimate is compared with exact numerical
calculation in  Fig.~\ref{fig:plate_joule}. From here it follows that the
Joule heat 
\begin{equation}
Q_J=\int_0^L dh \rho_J(h)
\approx \frac{1}{2}\sigma B_0^2 DL,
%+\frac{\sigma' B_0^2 D\lambda}{4}\ln(L/h_0).
\label{eq:nucl_joule}
\end{equation}
where we again take $\tanh(h/\lambda)=1$.  Moreover, we have assumed
that $h_0\ll L$.  The Joule heat is seen to be extensive, thereby
removing the unphysical features of the homogeneous transition
process.

It is instructive to study also the conservation of energy during the
nucleus growth. Starting with
Eq.~\eqref{eq:energy_conservation_general} and taking the integral
$\int_0^{L+\delta} dx$, we obtain a time-resolved analog
of~\eqref{eq:energy_ns}:
\begin{equation}
\frac{\partial U}{\partial t}+S(t)=P_B(t)-P_{\rm coil}(t),
\label{eq:energy_nucleation_plate}
\end{equation}
where $P_{\rm coil}(t)=B_0E(L,t)/\mu_0$ is the power delivered to the
coils and $P_B(t)$, defined by~\eqref{eq:W_B}, is the power which
needs to be supplied to the plate in order to expel the magnetic
field.

Plugging into this equation the estimates of the
fields~(\ref{eq:plate_nucleus_B},\ref{eq:plate_nucleus_E}) we find
\begin{equation}
P_B(t)=\frac{1}{2\mu_0\lambda^2}A^2(h,t)v
=\frac{B_0^2v}{2\mu_0}\tanh^2(h/\lambda).
\label{eq:pb_nucleation}
\end{equation}
Similarly, we find $P_{\rm coil}(t)=2P_B(t)$. Moreover, neglecting
terms proportional to $c^{-2}$, we also find that $\partial U/\partial
t=-P_B$. Since the Poynting vector $S(t)$ is proportional to $c^{-1}$,
the condition for energy
conservation~\eqref{eq:energy_nucleation_plate} is thus seen to be
satisfied by the approximate solution to order $c^0$.

As explained in III.B, the power $P_B$ which needs to be supplied to
the plate has to come from the free energy difference between the
normal and superconducting states. Since the density of this energy
difference is $B_c^2/(2\mu_0)$, the power (per unit area of the plate)
generated by an interface moving at speed $v$ is $P_{\Delta}(t)=
B_c^2v(t)/(2\mu_0)$. Note that, if $B_0=B_c$, the power
$P_{\Delta}$ is just enough to supply the needed power $P_B$. The
equality $P_{\Delta}=P_B$ is valid for all times except for the very
first stages of the nucleation when $h\lesssim\lambda$. It should be
pointed out that, at these times, our approximations for the fields
don't work either.

The above argument allows us to determine the functional form of
$h=h(t)$.  In fact, since the nucleation process takes place in an
external field which is smaller than the critical field, $B_0<B_c$, we
find that $P_{\Delta}>P_B$ and the surplus energy has to be removed
from the electrons. Let us assume that this energy is pumped in the
form of Joule heat $P_J$ into the lattice degrees of freedom,
\begin{equation}
P_{\Delta}=P_B+P_J.
\label{eq:powers_ns}
\end{equation} 
Plugging into this equation the expressions for $P_{\Delta}$, $P_B$,
and $P_J$ and assuming that $h>\lambda$, one can show that the front
velocity has to satisfy an equation of the same form
as~\eqref{eq:front_velocity} with the diffusion constant given by
\begin{equation}
D=\frac{|B_c^2-B_0^2|}{B_0^2}\frac{1}{\mu_0\sigma},
\label{eq:diff_constant}
\end{equation}
in agreement with \cite{Livingston69}. Making use of this result in
the expression for the Joule heat~\eqref{eq:nucl_joule} we thus find
\begin{equation}
Q_J\approx \frac{|B_c^2-B_0^2| L}{2\mu_0}.
\label{eq:joule_heat_nucleation}
\end{equation}

One observes that $Q_J$ does not depend on the normal-state
conductivity $\sigma$.  That this must be so follows from the
following argument. The total dissipated energy $W_{\rm diss}$ has to
be equal to the difference between the total generated energy $\Delta
F=\int_0^{t_{\rm max}} dt P_{\Delta}(t)=B_c^2 L/(2\mu_0)$ and the
energy required to expel the field, $W_B=\int_0^{t_{\rm max}} dt
P_B(t)\approx B_0^2 L/(2\mu_0)$. Therefore $W_{\rm diss}=\Delta F-W_B$
is given by the same expression as~\eqref{eq:joule_heat_nucleation}.

From~\eqref{eq:powers_ns} it follows that, since $P_\Delta$ is
associated with emission of latent heat, a total heat of
\begin{equation}
Q_{ns}=T\Delta S+Q_J
\label{eq:total_heat_ns}
\end{equation}
is released to the reservoir when a normal metal transforms
isothermally into a superconductor. One should note that, for
processes driven by a small ``undercooling'' $B_c-B_0$, the Joule heat
$Q_J$ is just a small fraction of the latent heat $T \Delta S$ which
is of the order of $\Delta F$, at least for processes which occur at
temperatures $T$ not too close to $0$ or $T_c$ \cite{Hirsch_heat}.

Before concluding, a comment is in place here. Namely, in the studied
nucleation process the velocity of the front increases as a function
of time. It has been argued that this implies that a planar front
should be unstable in such a case. Moreover, experimental studies
suggest that the nucleation process starts at the sample surface and
not in its interior \cite{Livingston69}. We have therefore solved the
equations of the London electrodynamics with a planar nucleation front
moving from the plate surface to the plate center. While we find that
the Meissner effect does take place also in this scenario, the Joule
heat is more than extensive in this case. We believe that this
indicates that a completely satisfactory description of the transition
process from the normal metal to a superconductor can not be treated
as a one-dimensional problem and it therefore is beyond the scope of
the present paper. However, in the next section we will show how to
circumvent this complication.

\subsection{Nucleation of normal metal in a superconductor}
Let us study the field-first protocol reverted in time: in the initial
state, the plate is fully superconducting and the applied field $B_0$
is larger than the critical field $B_c$, so that the thermodynamically
stable state of the plate is a normal metal. Thus the plate should
transform into a normal metal. Experimentally it is known that the
normal phase nucleates close to the plate surface and the nucleation
front $h(t)$ moves towards the plate center \cite{Livingston69}.

Later we will show that, in agreement with \cite{Livingston69}, the
differential equation for the velocity of the front in this case is
$\dot{h}=-D/(2(L-h))$. The solution to this equation which satisfies
the initial condition $h(0)=L$ reads $h(t)=L-\sqrt{Dt}$, implying that
$v(t)=-D/(2\sqrt{Dt})$. This means that the absolute value of the
front velocity decreases with time and the planar interface may be
stable in this case. Thus, in the present case, the nucleation
protocol described by the ``condensate fraction''
$f(x,t)=\theta(h-|x|)$ is well justified, both experimentally and
theoretically \cite{Livingston69}. In order to prevent the unphysical
divergence of the velocity at $t=0$, similarly as in III.C we assume
that $v(t)=-v_{\rm max}$ in the initial stages of the nucleation
process when $h(t)>L-h_0$.

We have numerically solved the equations of the London electrodynamics
with the initial condition~\eqref{eq:meissner_plate} for the magnetic
field and the initial electric field $E(x,0)=0$. The results (not
shown) are qualitatively very similar to those shown in
Fig.~\ref{fig:plate_nucleus_ns}, but in reverted time order. The
magnetic and electric fields are again reasonably well described by
Eqs.~(\ref{eq:plate_nucleus_B},\ref{eq:plate_nucleus_E}). Note,
however, that in the present case $v(t)<0$ and therefore the electric
field in the right half of the plate is negative.  Thus, looking at
the acceleration equation~\eqref{eq:acceleration}, one might expect
that the absolute value of the negative supercurrent increases,
opposing the flux entering the sample. However, similarly as in III.C,
much more important is the second term in~\eqref{eq:acceleration},
which leads to a decrease of the absolute value of the supercurrent as
a function of time, since $\partial f/\partial t<0$ and therefore
$\partial j_s/\partial t>0$.

As regards the energetics, analogously to III.C, the transformation
process generates, due to the inflow of the magnetic field, power
$P_B(t)=B_0^2|v(t)|/(2\mu_0)$. Part of this power is used to deliver
the required condensation energy, $P_\Delta=B_c^2|v(t)|/(2\mu_0)$, and
the rest has to be dissipated.  Let us assume again that the energy is
dissipated as the Joule heat $P_J$,
\begin{equation}
P_B=P_{\Delta}+P_J.
\label{eq:powers_sn} 
\end{equation}
Plugging into this equation the expressions for $P_{\Delta}$, $P_B$,
and $P_J$ and assuming that $h>\lambda$ one can again show, exactly as
in III.C, that the front velocity satisfies the equation
$\dot{h}=-D/(2(L-h))$ with the diffusion constant given
by~\eqref{eq:diff_constant}.  Making use of this result we find that
the Joule heat is again given by~\eqref{eq:joule_heat_nucleation},
see also Fig.~\ref{fig:plate_nucleus_polia}.

Hirsch has posed also the following question: what is the force which
makes the nucleation front to move?  In order to answer this question,
consider random fluctuations of the front position. For the sake of
concreteness, let us assume that $B_0>B_c$.  Our analysis shows that,
in this case, the probability that the front moves to the plate center
is larger than that it moves to the plate surface, because in the
former process the energy decreases, while in the latter one it
increases. As a result, it is the normal phase
which grows.  Obviously, for the same reason it is the
superconducting phase which grows in the case  when $B_0<B_c$.

From~\eqref{eq:powers_sn} it follows that, since $P_\Delta$ is
associated with absorption of latent heat, a total heat of
\begin{equation}
Q_{sn}=T\Delta S-Q_J
\label{eq:total_heat_sn}
\end{equation} 
needs to be supplied to the plate from the reservoir in order
to transform a superconductor into a normal metal.

In a completely reversible transition, the heat $Q_{ns}$ has to be
equal to $Q_{sn}$. This happens when the Joule heat vanishes,
$Q_J=0$. We have shown that, if the applied field $B_0$ differs from
the critical field $B_c$, the transition is never strictly reversible.
However, it can be made as close to reversible as one wishes by
decreasing the value of $|B_0-B_c|$.

Very recently, Hirsch has suggested that standard theory of
superconductivity has the following problem \cite{Hirsch25}: The
energy $W_B=B_0^2L/(2\mu_0)$ generated in the process when the
superconductor slowly transitions into the normal state is equal to
the energy of the stopped supercurrents, and the conventional theory
provides no mechanism for converting it into the free energy
difference between normal and superconducting states without
dissipation.

From our Eq.~\eqref{eq:powers_sn} it follows that $W_B=\Delta F+Q_J$
and only a minor part of $W_B$ is produced as Joule heat.  Essentially
the whole energy $W_B$ has to be used to (reversibly) break the Cooper
pairs into normal electrons. Let us note that, in order to break a
Cooper pair, also entropy has to be supplied to the system. The
necessary entropy is provided by absorbing the heat $Q_{sn}$. The
total energy increase of the plate caused by disappearance of the
condensate therefore is $W_B+Q_{sn}=\Delta{\cal E}$, as required by
thermodynamics. Although our formalism does not allow us to study the
kinetics of the conversion process between the superfluid and normal
electrons, we conclude that Hirsch's problem (iv) from the
Introduction is not incompatible with the London electrodynamics
\cite{Hirsch_entropy}.

%%%%%%%%%%%%%%%%%%%%%%%%%%%%%%%%%%%%%%%%%%%%%%%%%%%%%%%%%%%%%%%%%%%%%%%%%%%
\section{Dynamics of the Becker-London experiment}
In this Section we analyze the simplest possible geometry of the
Becker-London experiment: we assume that the rotating body has the
shape of a long cylinder with radius $R\gg\lambda$. We work in an
inertial laboratory frame and we assume that the cylinder rotates
along its axis with angular velocity $\omega$. We make use of the
cylindrical coordinates, and, therefore,
$\bm{\omega}=(0,0,\omega)$. The rotating ionic charge with density
$\rho$ produces "external" current density ${\bf j}_{\rm
  ext}=\rho\bm{\omega}\times{\bf r}$, which in turn generates magnetic
field and possibly also electronic currents.  Obviously, only the
$\varphi$-component of the current ${\bf j}_{\rm ext}$ is
non-vanishing, ${\bf j}_{\rm ext}=(0,j_{\rm ext}(r),0)$, and $j_{\rm
  ext}=\rho\omega r$ depends only on $r$.
  
In the chosen geometry, also the electronic current, the electric
field, and the vector potential in the London gauge are non-vanishing
only in the $\varphi$-direction. Moreover, these fields depend only on
the $r$-coordinate.  Similarly, the magnetic field turns out to be
given by ${\bf B}=(0,0,B(r))$. The Maxwell equations therefore reduce
to
\begin{eqnarray}
\frac{1}{c^2}\frac{\partial E}{\partial t}&=&
-\frac{\partial B}{\partial r}-\mu_0(j_{\rm ext}+j_s),
\label{eq:Maxwell_rot_E}
\\
\frac{\partial B}{\partial t}&=&
-\frac{1}{r}\frac{\partial }{\partial r}(rE),
\label{eq:Maxwell_rot_B}
\\
\frac{\partial A}{\partial t}&=&-E.
\label{eq:Maxwell_rot_A}
\end{eqnarray}
Once the initial values of the fields
$B(r,0)=r^{-1}\partial(rA)/\partial r$ and $E(r,0)$ are known, the
future temporal evolution of the fields $E$, $B$, and $A$ can be
calculated from these equations, provided also an equation for the
currents $\mu_0(j_{\rm ext}+j_s)$ is specified.

Our goal is to calculate the magnitude of the magnetic field which
develops in a rotating superconductor described by the London
electrodynamics. As discussed in the Introduction, we consider two
different experimental protocols: the cool-first protocol and the
rotate-first protocol.  The question is whether both protocols lead to
the same final state of the cylinder.

\subsection{Setting a superconductor into a rotating state}
In this case the cylinder is at rest at times $t<0$, and as a result
also $j_{\rm ext}=0$ and no electromagnetic fields are present at the
initial time $t=0$, i.e. $B(r,0)=0$ and $E(r,0)=0$. At times $t>0$ the
cylinder starts to rotate and, for definiteness, we assume that the
angular frequency changes as $\omega g(t)$, where
$g(t)=\tanh(t/\tau)$. Therefore the ``external'' current due to the
motion of the nuclei inside the cylinder is $j_{\rm
  ext}(r,t)=\rho\omega r g(t)$. Since all (conduction) electrons are
supposed to form the condensate, the London equation~\eqref{eq:London}
simplifies to $\mu_0 j_s=-A/\lambda^2$, and therefore the total
current inside the cylinder is given by
\begin{equation}
\mu_0(j_{\rm ext}+j_s)=\frac{B_0}{2\lambda^2} r g(t)-\frac{1}{\lambda^2}A,
\label{eq:current_rot_prot1}
\end{equation}
where we have introduced the notation $B_0=2\mu_0\rho\omega\lambda^2$.

Although
Eqs.~(\ref{eq:Maxwell_rot_E},\ref{eq:Maxwell_rot_B},\ref{eq:Maxwell_rot_A})
supplemented by~\eqref{eq:current_rot_prot1} are well suited for the
numerical solution of the initial-value problem at hand, it is useful
to reformulate them also only in terms of the Proca-like
equation~\eqref{eq:Klein_Gordon_general}, which simplifies to
\begin{equation}
\left[\frac{1}{c^2}\frac{\partial^2}{\partial t^2}+\frac{1}{\lambda^2}
-\bigtriangleup\right]B
=\frac{B_{\rm eff}(t)}{\lambda^2},
\label{eq:Klein_Gordon_cool}
\end{equation}
where the time-dependent effective magnetic field $B_{\rm eff}(t)$ is
given by an equation formally identical with~\eqref{eq:B_eff}.

In the limit of long times $t\gg\tau$, one expects that the solution
to this equation becomes time-independent and reads
\begin{equation}
B(r)=B_0\left[1-\frac{I_0(r/\lambda)}{I_0(R/\lambda)}\right],
\label{eq:B_rotating}
\end{equation}
where $I_0(x)$ is the modified Bessel function. Here we have used the
fact that the boundary condition for a static field requires that
$B(R)=0$.

Thus, in the limit $R\gg\lambda$, the field generated deep inside a
rotating superconductor is $B_0$.  Following
Hirsch~\cite{Hirsch19annalen}, we make use of the estimate
$\mu_0\lambda^2=m/(\rho e)$ where $m$ is the free electron mass, and
we find that $B_0=2m\omega/e$. It is worth pointing out that in a
series of theoretical papers it has been argued that using the free
electron mass in the formula for the Becker-London field $B_0$ is in
fact correct, although the penetration depth actually depends on the
band mass $m^\ast$ \cite{Darwin36,Geurst93,Lipavsky13}.  In passing,
let us also note that the expression $B_0=2m\omega/e$ is actually in
very good agreement with experimental data for a broad spectrum of
superconductors, see \cite{Tate89,Verheijen90,Sanzari96}.

\begin{figure}[t]
\includegraphics[width = 7. cm]{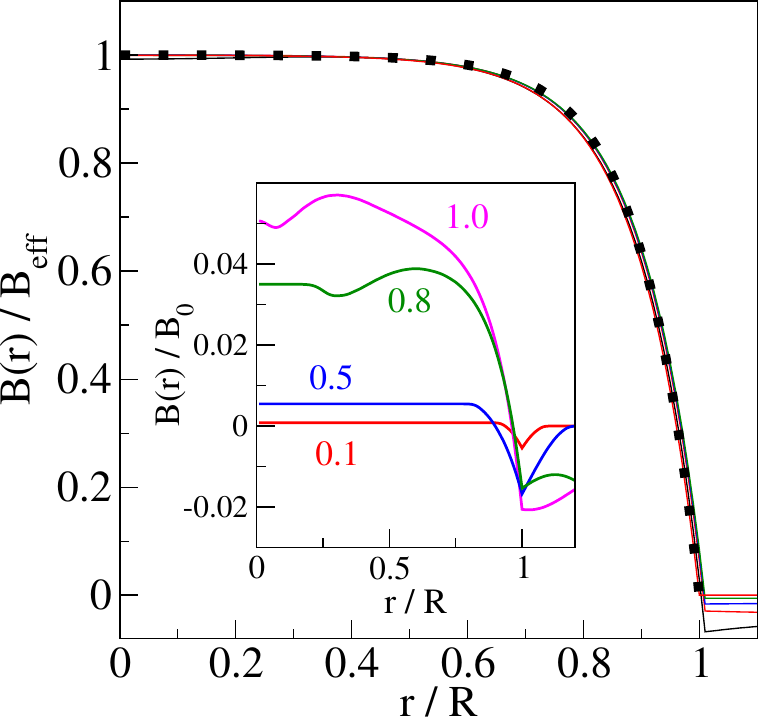}
\caption{Temporal evolution of the magnetic field profile in the
  rotating cylinder (for $\lambda/R=0.1$ and $c\tau/R=20$) for the
  cool-first protocol. Data for $ct/R=$5, 10, 15, 25, and 100 are
  shown in the main panel. The dots display Eq.~\eqref{eq:B_rotating}
  with $B_{\rm eff}=B_0$. The inset shows a detailed view of $B(r,t)$
  for the indicated short times.}
\label{fig:rot_cool}
\end{figure}

Numerical results for the temporal evolution of the magnetic field
profiles inside the cylinder are plotted in
Fig.~\ref{fig:rot_cool}. We find that at short times $t<R/c$ and
distances larger than $ct$ from the cylinder rim, the magnetic field
does not depend on the spatial coordinate $r$ and is perfectly flat.

In the initial stage of the Becker-London process, i.e. for $t\lesssim
t_{\rm initial}=\sqrt{6}\lambda/c$, the electromagnetic field in the central
region of the cylinder is given by
\begin{equation}
B(r,t)\approx \frac{c^2t^3 B_0}{6\lambda^2\tau},
\qquad
E(r,t)\approx -\frac{rc^2 t^2 B_0}{4\lambda^2\tau}.
\label{eq:rotating_fields}
\end{equation}
Extracting from here the vector potential we also find
\begin{equation}
\mu_0 j_s=-\frac{r c^2 t^3 B_0}{12\lambda^4\tau}.
\end{equation}
Note that, in agreement with the predictions of Hirsch, the
acceleration of the supercurrent $j_s$ is described
by~\eqref{eq:acceleration} with a time-independent superconducting
fraction.

As also shown in Fig.~\ref{fig:rot_cool}, for $t\sim R/c$ the magnetic
field profile has a quite complicated shape, but for later times
$t\gtrsim R/c$ it is well described by~\eqref{eq:B_rotating}, if $B_0$
is replaced by the effective time-dependent field $B_{\rm eff}$ given
by~\eqref{eq:B_eff}. One observes readily that the functional
form~\eqref{eq:B_rotating} solves~\eqref{eq:Klein_Gordon_cool} if we
neglect the time derivative, or, in other words, if the Maxwell
displacement current is ignored.

\subsection{Cooling a rotating normal metal}
In this case, although the cylinder is rotating, the total
``external'' current flowing at times $t<0$ inside the cylinder is
$j_{\rm ext}=0$. The reason is the following: since for $t<0$ the
cylinder is in the non-superconducting state, the friction between the
electrons and the ions is finite, and, therefore, the electrons
ultimately rotate at the same angular frequency $\omega$ as the ions,
generating the current density $j_{\rm el}=-\rho\omega r$. But $j_{\rm
  ext}=j_{\rm nucl}+j_{\rm el}$, and since $j_{\rm nucl}=+\rho\omega
r$, we find therefore that $j_{\rm ext}=0$.  In other words, if the
cylinder is in the normal state, it can be considered as an uncharged
structureless object which rotates as a whole. Therefore at time $t=0$
we assume that $B(r,0)=0$ and $E(r,0)=0$.

Things change for times $t>0$, when we start to transform the normal
metal into a superconductor. Let us assume that, at time $t$, the
superconducting fraction is $f(t)=\tanh(t/\tau)$. Adopting the
standard two-fluid description, the fraction of normal electrons
equals $1-f$ and, therefore, the normal-electron part of the
``external'' current is $j_{\rm el}=-(1-f)\rho\omega r$. This implies
that the total external current density is $j_{\rm ext}=j_{\rm
  nucl}+j_{\rm el}=f(t)\rho\omega r$. Adding the external current with
the supercurrent, we thus finally find
\begin{equation}
\mu_0(j_{\rm ext}+j_s)=\frac{f(t)}{\lambda^2}
\left[\frac{B_0r}{2} -A\right].
\label{eq:current_rot_prot2}
\end{equation}
 
Similarly as in the previous subsection,
Eqs.~(\ref{eq:Maxwell_rot_E},\ref{eq:Maxwell_rot_B},\ref{eq:Maxwell_rot_A})
supplemented this time by~\eqref{eq:current_rot_prot2} can be again
written as a single ``driven'' Proca-like equation,
\begin{equation}
  \left[\frac{1}{c^2}\frac{\partial^2}{\partial t^2}
    +\frac{1}{\lambda_{\rm eff}(t)^2}
-\bigtriangleup\right]B
=\frac{B_0}{\lambda_{\rm eff}(t)^2},
\label{eq:Klein_Gordon_rot}
\end{equation}
where the time-dependent penetration depth $\lambda_{\rm eff}(t)$ is
given by~\eqref{eq:lambda_eff}.

Experiments tell us that, in the long-time limit, the magnetic field
inside the cylinder is again given by~\eqref{eq:B_rotating}, as in the
previous subsection \cite{Hendricks71}. Our goal here is to explain
this result within the London electrodynamics.

Let us first address the claim of Hirsch that, according to
conventional theory, the superconducting electrons should remain
rotating together with the ions in the rotate-first protocol,
$j_s+j_{\rm ext}=0$, and therefore no magnetic field should be
generated. This argument is incorrect: In fact,
from~\eqref{eq:current_rot_prot2} it follows that $j_s+j_{\rm ext}=0$
implies $A=B_0r/2$, and therefore the field inside the cylinder should
be $B=B_0$ and not zero.  Moreover, although this looks like an
acceptable solution to
Eqs.~(\ref{eq:Maxwell_rot_E},\ref{eq:Maxwell_rot_B},\ref{eq:Maxwell_rot_A},\ref{eq:current_rot_prot2}),
it is not correct, because it respects neither the boundary condition
$B(R,\infty)=0$, nor the initial condition $B(r,0)=0$.

\begin{figure}[t]
\includegraphics[width = 7. cm]{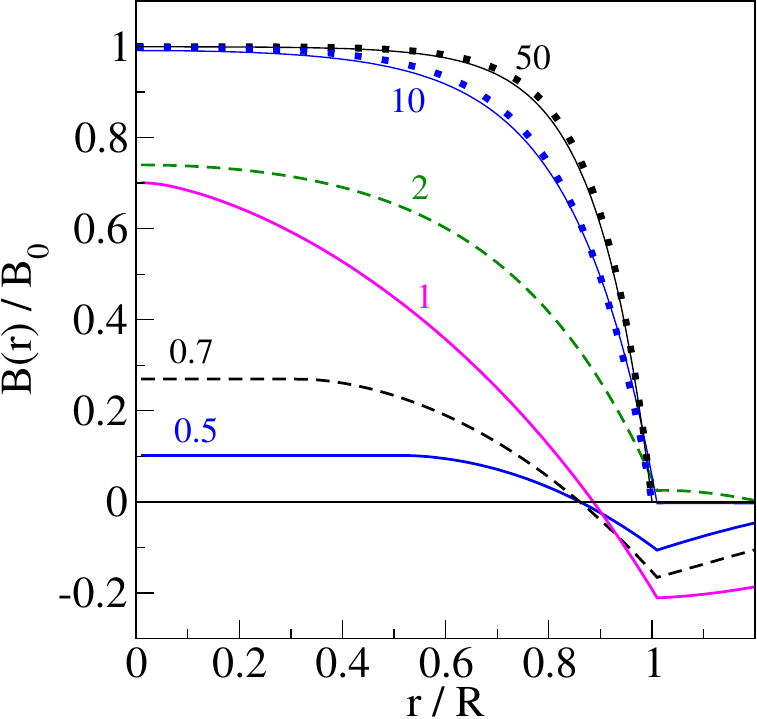}
\caption{Temporal evolution of the magnetic field profile in the
  rotating cylinder (for $\lambda/R=0.1$ and $c\tau/R=20$) for the
  rotate-first protocol. The numbers indicate time in units of
  $R/c$. The dots correspond to Eq.~\eqref{eq:B_rotating} with
  $\lambda$ replaced by $\lambda_{\rm eff}(t)$.}
\label{fig:rot_rot}
\end{figure}

So why does the process of $B$-field generation start?  In the initial
stage of the process the magnetic field vanishes, and therefore also
the vector potential $A=0$. As a result, from the London
equation~\eqref{eq:London} it follows that the supercurrent does not
move, $j_s=0$.  Thus the external current $j_{\rm ext}$ is not
compensated by $j_s$ and a finite net electric current flows in the
system, which in turn generates the magnetic field. When compared with
the field-first protocol for the Meissner effect, where we had to ask:
``what drives the supercurrent?'', in this case we ask instead: ``what
keeps the supercurrent at rest?'' The answer in both cases is the
same: it is the instantaneous magnitude of the vector potential.

Numerically calculated temporal evolution of the magnetic field
profiles inside the cylinder for the rotation-first protocol is
plotted in Fig.~\ref{fig:rot_rot}.  We find again that at times
$t<R/c$ and distances larger than $ct$ from the cylinder rim, the
magnetic field does not depend on the spatial coordinate $r$, i.e. it
is flat.

At the shortest times $t\lesssim t_{\rm
  initial}=(6\tau\lambda^2/c^2)^{1/3}$, the electromagnetic field in
the center of the cylinder is still given
by~\eqref{eq:rotating_fields}, and therefore the supercurrent is given
by
\begin{equation}
\mu_0 j_s=-\frac{r c^2 t^4 B_0}{12\lambda^4\tau^2}.
\end{equation}
The results for $j_s$ and $E$ are again in agreement
with~\eqref{eq:acceleration}, but this time with a superconducting
fraction $f\approx t/\tau$.

At intermediate times $t\sim R/c$, the magnetic field profile is again
complicated, see Fig.~\ref{fig:rot_rot}. However, for later times
$t\gtrsim R/c$ it is well described by~\eqref{eq:B_rotating}, if
$\lambda$ is replaced by $\lambda_{\rm eff}(t)$. Note that the
functional form~\eqref{eq:B_rotating}
solves~\eqref{eq:Klein_Gordon_rot}, if we neglect the Maxwell
displacement current.

\begin{figure}[t]
\includegraphics[width = 7. cm]{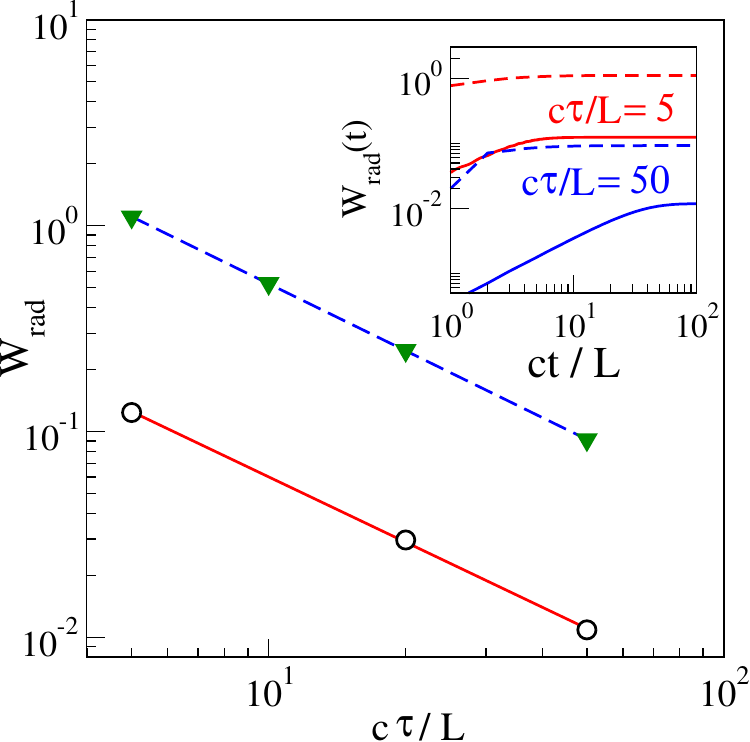}
\caption{Total energy $W_{\rm rad}$ (in units of $B_0^2R/\mu_0$)
  radiated by the unit length of the rotating cylinder (for
  $\lambda/R=0.1$) in the cool-first protocol (full lines) and
  rotate-first protocol (dashed lines), as a function of the switch-on
  duration $\tau$. The inset shows the temporal evolution of $W_{\rm
    rad}(t)$ (see text) for two different values of $\tau$.}
\label{fig:rot_radiation}
\end{figure}

Let us remark that in both protocols, the cylinders emit radiation
while the magnetic field inside them changes.  In
Fig.~\ref{fig:rot_radiation} we plot the energy emitted by unit length
of the rotating cylinder up to time $t$, $W_{\rm rad}(t)=2\pi R
\int_0^t dt' S(R,t')$, as well as the total emitted energy $W_{\rm
  rad}= W_{\rm rad}(\infty)$ for both protocols. As expected, $W_{\rm
  rad}$ scales as $\tau^{-1}$ and the radiation lasts approximately
for time $\tau$. Note that $W_{\rm rad}$ is consistently larger for
the rotate-first protocol than for the cool-first protocol. We have
also checked that energy is conserved in both protocols. The relevant
formulae for various types of energy are given, for the sake of
completeness, in Appendix~B.

Before concluding we would like to point out that, when treating the
rotate-first protocol, we have assumed that the condensate acquires
phase coherence uniformly throughout the cylinder. This is certainly
not realistic. In Appendix~C we therefore consider a more realistic
scenario where superconductivity nucleates along the cylinder axis and
afterwards the nucleus grows radially. We show there that also in this
scenario the ultimate distribution of the magnetic field inside the
cylinder is again described by Eq.~\eqref{eq:B_rotating}.

Yet another point worth mentioning follows from the fact that, for
experimentally achievable rotation frequencies $\omega$, the magnetic
field $B_0$ is typically very small compared with the critical field
at zero temperature $B_c$. Therefore the transition from the normal to
the superconducting state in the rotate-first protocol is nearly of
second order even in a type-I superconductor. Nevertheless, similarly
as in Section~III.B, we have assumed that the condensate formation is
not affected by this fact. This assumption does agree well with
experimental results in very clean tin cylinders
\cite{Hendricks71}. However, in cylinders where impurities do play a
role the experimental results are different \cite{Hendricks71}. A
serious analysis of that problem is beyond the scope of the present
paper, but in Appendix~D we present some speculations on this issue.

%%%%%%%%%%%%%%%%%%%%%%%%%%%%%%%%%%%%%%%%%%%%%%%%%%%%%%%%%%%%%%%%%%%%%%%%%%%
\section{Discussion}

\subsection{Validity of the London equation}
In this manuscript we have assumed that the London
equation~\eqref{eq:London} applies also in cases with a time- and
space-dependent superconducting fraction $f({\bf x},t)$ and we have
shown that, once this assumption is made, the dynamics of the Meissner
and the Becker-London effects can be straightforwardly understood.
Before concluding we present some arguments supporting this
assumption.

In textbooks, the {\it local} relation between the current and the
vector potential, Eq.~\eqref{eq:London}, is usually justified in the
homogeneous case and in thermal equilibrium. Therefore we believe that
Eq.~\eqref{eq:London} should be still at least approximately valid
also for a condensate which changes sufficiently slowly on the scale
of the Cooper-pair size and for processes slow enough so that the
system is always close to thermal equilibrium.

What does this imply for the field-first protocol of the Meissner
effect?  When the nucleation front moves across the sample, the above
conditions for the validity of Eq.~\eqref{eq:London} are probably not
satisfied right at the front position. One can, however, define a
second front, lagging slightly behind the nucleation front, where
thermal equilibrium is already established - and where
Eq.~\eqref{eq:London} should be applicable. 

Yet another argument is based on the time-dependent Ginzburg-Landau
(TDGL) equations: they do lead to Eq.~\eqref{eq:London}, see, e.g.,
\cite{Tinkham04}. Although the TDGL theory is justified in only a very
limited range of situations, we believe that a functional form
resembling Eq.~\eqref{eq:London} is a natural consequence of the
quantum-mechanical formula for the current carried by a macroscopic
wave-function. Of course, more complete formulas might include
modifications such as Pippard's non-local kernel, a similar
non-locality in time, etc.

Finally, Eq.~\eqref{eq:London} may be justified a posteriori by the
acceleration equation~\eqref{eq:acceleration}, which has a very
natural interpretation. In fact, according to
Eq.~\eqref{eq:acceleration}, the supercurrent may change for two
reasons: First, in an electric field, the Cooper pairs accelerate,
thereby changing the supercurrent. Second, exchange of electrons
between the normal fluid and the moving condensate also changes the
supercurrent.

\subsection{Analogy with superfluid $^4$He}
Another question asked by Hirsch is: what 'force' imparts momentum
to the normal electrons when joining a moving condensate?

An analogy to superfluid $^4$He helps to understand what is going
on. Experiments on persistent flow in an annular container show that,
when the container is cooled from an initial state which carries a
finite superflow, those $^4$He atoms which were originally in the
normal fluid enter the condensate, increasing the angular momentum of
the condensate. Similarly, when the container is heated, the angular
momentum of the condensate decreases \cite{Reppy65}.

The standard interpretation of these findings which we adopt here is
that, once the superfluid flow is established, it is the phase (or
velocity) field of the condensate which is frozen in at the moment
when the symmetry is spontaneously broken \cite{Goodstein85}. In such
an externally prescribed velocity field, the free energy at any given
temperature is minimized by an optimal, temperature-dependent,
condensate density.

So the 'force' imparting momentum to the helium atoms when joining (or
leaving) the condensate as the temperature changes is of thermodynamic
origin: we are dealing with a dynamic equilibrium between the normal
fluid and the condensate.  The situation is in fact quite similar to
that of phase equilibrium in chemical reactions: when temperature
changes, also the concentration of the reagents changes.

In the case of a superconductor, an analogous 'force' is responsible
for the exchange of pairs electrons between the normal fluid and the
condensate. The main difference between equilibrium in chemical
reactions on one hand, and equilibrium in rotating superfluids and the
Meissner effect on the other hand, lies in the fact that in the latter
two examples the changing number of particles in the condensate
implies a change of a conserved quantity, namely momentum.

In the $^4$He experiment \cite{Reppy65}, it is the walls of the
container which play the role of the momentum sink which is necessary
for momentum conservation. Similarly, in the Meissner effect the role
of the momentum sink is played by the ionic lattice. In both
superfluids and superconductors, the gross picture is the same:
momentum is exchanged between the condensate and the normal fluid (by
exchange of particles) and between the normal fluid and the momentum
sink (by scattering processes).

Experiments tell us that both, the persistent flow of helium in an
annular container and the Meissner screening current in the
field-first protocol, change reversibly for sufficiently slow
processes. This means that the exchange of momentum between the normal
fluid and the momentum sink in none of these situations spoils
reversibility. 

It is fair to say that a detailed microscopic justification of the
observed reversibility within conventional theory still needs to be
worked out. We believe that the analogous behavior of uncharged
superfluids and charged superconductors strongly calls for a common
explanation in both cases.

%%%%%%%%%%%%%%%%%%%%%%%%%%%%%%%%%%%%%%%%%%%%%%%%%%%%%%%%%%%%%%%%%%%%%%%%%%%
\section{Conclusions}
In this paper we have shown that when the Maxwell equations are
supplemented by the simplest constitutive equation of the conventional
theory of superconductivity, namely the London
equation~\eqref{eq:London}, the problems identified by Hirsch in the
conventional analysis of the dynamics of the Meissner and the
Becker-London effects do not exist.

The crucial difference between our approach and Hirsch's analysis is
rooted in the formulation of the constitutive equations of the
conventional theory. Hirsch formulates them in terms of the fields
${\bf E}$ and ${\bf B}$, whereas we take for the London
equation~Eq.\eqref{eq:London}, which expresses the supercurrent ${\bf
  j}_s$ in terms of the vector potential ${\bf A}$ in the London
gauge.  This innocent-looking difference in formulation leads in our
version of the theory to the appearance of new terms in
Eqs.~(\ref{eq:screening},\ref{eq:acceleration}) in spatially and
temporally varying superconductors, which are not present in Hirsch's
formulation.

Another important point of our analysis is that, when looking at the
dynamics of the transformation process from a metal to a
superconductor or vice versa, one obviously can not neglect the time
derivative of the superconducting fraction $\partial f/\partial t$.

These observations lead to a simple qualitative explanation of the
points (ii) and (iii) raised by Hirsch: Due to the presence of the
dominant new term in the acceleration
equation~\eqref{eq:acceleration}, the supercurrent either accelerates
or decelerates in the direction of the vector potential, depending on
the sign of $\partial f/\partial t$.  Moreover, Hirsch's point (i) is
also answered, since the magnitude of the supercurrent at the very
first moment of the transformation process is controlled by the
instantaneous value of the vector potential at that time.

From what has been said so far it follows that it is crucial to
justify the use of the vector potential in the London
equation~\eqref{eq:London}. We have two arguments in favor of this
(standard) choice: First,~\eqref{eq:London} can be considered as the
limiting case of the Ginzburg-Landau formula for the supercurrent
carried by a condensate with a fixed amplitude $|\psi|$. Although the
Ginzburg-Landau theory is also phenomenological
\cite{deGennes66,Tinkham04}, it does reflect the conventional view of
the superconductor as a macroscopic condensate. Since this theory is
of inherently quantum nature, by necessity it expresses the
supercurrent in terms of the vector potential.  Second, also the
Aharonov-Bohm effect can be described in the most natural way in the
language of electromagnetic potentials instead of fields
\cite{Aharonov59}.

Regarding the question about the validity of our conclusions, one may
adopt also a more pragmatic approach: excluding the vector potential
and the electric field, one can model the processes under question by
Proca-like equations for the magnetic field only
\cite{note_KG}. Simple examples of such equations are given by
Eqs.~(\ref{eq:Klein_Gordon_cool},\ref{eq:Klein_Gordon_rot}). A
brute-force solution of such equations demonstrates that the
conventional theory of superconductivity does describe the dynamics of
the Meissner and Becker-London effects correctly, irrespective of the
chosen protocol. The magnetic field profiles in both cases are
described at long times by~\eqref{eq:meissner_plate}
and~\eqref{eq:B_rotating}, respectively. The difference between the
studied protocols enters
Eqs.~(\ref{eq:meissner_plate},\ref{eq:B_rotating}) as follows: for the
cool-first protocol, the amplitude $B_0$ has to be replaced by the
effective field $B_{\rm eff}(t)$, whereas for the field-first
(rotate-first) protocol, it is the penetration depth which has to be
replaced by $\lambda_{\rm eff}(t)$.

As a by-product of addressing Hirsch's criticism of the conventional
theory of superconductivity, this paper shows that the dynamics of
processes between different magnetic states of superconductors is
surprisingly rich and interesting. In general, at the shortest time
scales one observes fast relativistic propagation of the fields,
followed by a slower relaxation towards the final states. The duration
of the latter stage is typically determined by the time needed to
change the external parameters which drive the process at
question. 

Obviously, time-dependent magnetic fields within the sample generate
electromagnetic pulses carrying part of the energy from the external
forces away from the sample.  This complicates the numerical analysis: for
instance in case of the Meissner effect one can not simply prescribe
the magnitude of the external magnetic field at the sample
surface. Instead, one has to explicitly include the external currents
in the coils which generate the field and use absorbing boundary
conditions behind the coils.

There is a bonus we receive after solving this complication, however:
we can calculate in detail the energy balance of the studied
processes, having explicit knowledge not only of the sample energy,
but also of the work done by the external currents, energy needed to
expel the magnetic field, as well as of the radiated energy. The
results which we obtain are in full accordance with the simpler
thermodynamic arguments \cite{deGennes66,Tinkham04}, and generalize
them to the case of finite sample dimensions. 

In particular, we were able to identify the role of the energy $W_B$
which is released when the superconductor changes to a normal metal in
a finite magnetic field. This energy is not equal to the Joule heat
produced by stopping the supercurrent. Instead, $W_B$ is essentially
equal to the free-energy difference between the normal and
superconducting states in absence of the magnetic field. Therefore, we
believe that Hirsch's point (iv) does not apply either. It is fair to
say, however, that a detailed microscopic calculation of the entropy
produced when the supercurrent stops is still missing.

We would like to close with the following remark. Throughout this
paper we have assumed that phase coherence is already established in
the superconductor, so that we could apply to it the London
equations. In case of the cool-first protocols, this is well
justified: the processes start from perfect equilibrium states of the
superconductor and switching on (small) perturbations in the form of
the field or rotation did not change this starting configuration
appreciably.

The situation changes, however, when dealing with the field-first or
rotate-first protocols: in this case, the formation of phase coherence
in the condensate is part of the process. This point has been
repeatedly stressed by Hirsch. So how does phase coherence appear?
This is obviously a very complicated problem, related to the
Kibble-Zurek mechanism \cite{Kibble76,Zurek85,Lee24}. Obviously, an
approach based on the London equations can not really address it.

In order to shed at least some light on these difficult questions, we
have assumed that a single phase-coherent nucleation center already exists in
the sample. Afterwards we have assumed that the phase-coherent region
grows as a function of time and we have tracked the development of
electromagnetic fields for an externally prescribed condensate
fraction $f(x,t)$. At least in case of the field-first protocol for
the Meissner effect, this approach seems to be well justified, since
we are dealing with a phase transition of first order in this case.

In Sections III.C and III.D we have studied in this way the Meissner
effect and, in Appendix~C, the Becker-London effect.  Our calculations
fully confirm the simple results found for homogeneously increasing
condensate fraction.  We have also checked (not shown) that this
conclusion is valid also for different models of spatially non-uniform
growth of the condensate fraction.

Finally, in Appendix~D we address the following related problem: If
the rotating cylinder is hollow, the phase of the condensate can be
described by an arbitrary integer winding number $n$. Obviously, in
the cool-first scenario the winding number is $n=0$, but in the
rotate-first scenario any starting winding number $n$ can be frozen-in
in the process of condensate formation. Hirsch argues that, within
conventional theory, there does not exist a dynamic mechanism by which
the superconductor picks the correct winding number $n$ at the moment
of condensate formation \cite{Hirsch19annalen}. In Appendix~D we
suggest that such a mechanism might exist.

\begin{acknowledgments}
This work was supported by the Slovak Research and Development Agency
under Contract No.~APVV-23-0515.
\end{acknowledgments}

%%%%%%%%%%%%%%%%%%%%%%%%%%%%%%%%%%%%%%%%%%%%%%%%%%%%%%%%%%%%%%%%%%%%%%%%%%%
\appendix
\section{Momentum conservation in the plate}
As is well known, the Maxwell equations imply that
\begin{equation}
\frac{\partial\pi_i}{\partial t}=
\sum_j\frac{\partial T_{ij}}{\partial x_j},
\label{eq:momentum_conservation}
\end{equation}
where $\bm{\pi}=\bm{\pi}_{\rm field}+\bm{\pi}_{\rm mech}$ is the
momentum density of the field + matter system, and $T_{ij}$ is the
Maxwell stress tensor \cite{Jackson99}. In the plate geometry $T_{ij}$
is diagonal and it depends only on the $x$ coordinate, implying that
only the $x$-component of the right-hand side is non-vanishing.  The
field contribution to the momentum density is proportional to the
Poynting vector, $\bm{\pi}_{\rm field}={\bf S}/c^2=(S/c^2,0,0)$, while
the mechanical momentum density satisfies Newton's equation of motion
$d \bm{\pi}_{\rm mech}/d t={\bf f}$, where ${\bf f}=\rho{\bf E}+{\bf
  j}\times {\bf B}$ is the Lorentz force density. Note also that in
our geometry $d\bm{\pi}_{\rm mech}/d t=\partial\bm{\pi}_{\rm
  mech}/\partial t$.

The total charge density of the combined system electrons + ions is
$\rho=0$. Therefore the total force ${\bf f}=(jB,0,0)$ acting on the
plate is given by the $x$-component
of~\eqref{eq:momentum_conservation},
\begin{equation}
jB=-\frac{1}{c^2}\frac{\partial S}{\partial t}
-\frac{\partial u}{\partial x}.
\end{equation}
In the large-$\tau$ case when the Poynting vector can be neglected,
the Lorentz force is entirely given by the Maxwell stress $\partial
u/\partial x$, as is well known.

On the other hand, the $y$ and $z$ components of the total momentum
density $\bm{\pi}_{\rm mech}=\bm{\pi}_s+\bm{\pi}_{\rm nucl}$ of the
combined system electrons + ions are trivially conserved. However,
this does not mean that the momentum density of the superfluid
$\bm{\pi}_s$ does not change in the $y$-direction.  In order to show
this, let us realize that the superfluid current density ${\bf j}_s$
and the superfluid momentum density $\bm{\pi}_s$ are related by
\begin{equation}
\bm{\pi}_s=-\frac{m}{e}{\bf j}_s,
\label{eq:superfluid_momentum}
\end{equation}
where $m$ is the free electron mass.  From~\eqref{eq:acceleration} it
therefore immediately follows that $\partial\bm{\pi}_s/\partial t$ is
nonzero. If we assume that the normal electrons are tightly
bound to the nuclei,  by conservation of the total momentum of the system
electrons + ions in the $y$ direction one arrives at
\begin{equation}
\frac{\partial \pi_{\rm nucl}}{\partial t}=
\frac{m}{e}\frac{\partial j_s}{\partial t}
=nf eE-\frac{\partial (nf)}{\partial t}p_s.
\label{eq:acceleration_ions2}
\end{equation}
In the second equality, we have used~\eqref{eq:acceleration}, where we
took $\lambda^{-2}=\mu_0 ne^2/m$.  Note that the electron mass does
not appear in the final form of~\eqref{eq:acceleration_ions2}. As
shown in the main text, this equation has a very natural
interpretation, strongly suggestive that~\eqref{eq:acceleration_ions2}
is more general than our justification.

\section{Conservation of energy in the Becker-London experiment}
The continuity equation for the total energy of the system charges +
field reads \cite{Jackson99}
\begin{equation}
\frac{\partial u}{\partial t}
  +\nabla\cdot{\bf S}+{\bf j}\cdot{\bf E}=0.
\end{equation}
In cylindrical coordinates ${\bf S}=(S(r),0,0)$ where $S=EB/\mu_0$,
and therefore $\nabla\cdot{\bf S}=r^{-1}\partial (rS)/\partial r$.
Similarly, ${\bf j}\cdot{\bf E}=(j_{\rm ext}+j_s)E$. Next we plug
these results into the continuity equation and take an integral
$\int_0^\infty dt \int_0^R d^2r$ of the resulting equation.  

For the cool-first protocol, we find in this way that
\begin{equation}
U(\infty)+W_{\rm rad}=U(0)+W,
\label{eq:rot_cool_energy}
\end{equation}
where 
\begin{equation}
U(t)=\frac{1}{2\mu_0}\int_0^R d^2r
\left[\frac{1}{c^2}E^2+B^2+\frac{f}{\lambda^2}A^2\right]
\label{eq:rot_sample_energy}
\end{equation}
is the energy (per unit length of the cylinder) of the electromagnetic
field and of the superconducting current inside the
cylinder. Obviously, in the cool-first protocol, we have
$f=1$. Moreover, in the initial state $U(0)=0$.

Equation~\eqref{eq:rot_cool_energy} can be interpreted as follows: the
initial energy of the cylinder $U(0)$, increased by the work $W$
supplied to the cylinder by outside forces causing the rotation,
transforms into the final energy of the cylinder plus the energy
$W_{\rm rad}$ emitted by the radiation pulse. The mechanical work
supplied from outside is given by
\begin{equation}
W=-\frac{B_0}{2\mu_0\lambda^2}
\int_0^\infty dt  \int_0^R d^2r r f(r,t) E(r,t).
\label{eq:rot_mechanical_work}
\end{equation}
Note, however, that~\eqref{eq:rot_mechanical_work} does not include
mechanical work which is needed to make the cylinder rotate.
Obviously, additional mechanical work has to be supplied in order to
set the nuclei into a rotating state.

Applying the same procedure, for the rotate-first protocol we
similarly obtain
\begin{equation}
U(\infty)+W_{\rm rad}=U(0)+W+W_B.
\label{eq:rot_rot_energy}
\end{equation}
In this case, the energy balance is slightly more involved
than~\eqref{eq:rot_cool_energy}: in addition to mechanical work $W$,
the sample also absorbs energy needed to generate the magnetic field,
\begin{equation}
W_B=\frac{1}{2\mu_0\lambda^2}
\int_0^\infty dt  \int_0^R d^2 r \frac{\partial f}{\partial t} A^2(r,t).
\label{eq:rot_thermal_work}
\end{equation}
This energy is supplied by the transformation of the normal to the
superconducting state.

\begin{figure}[b]
\includegraphics[width = 7. cm]{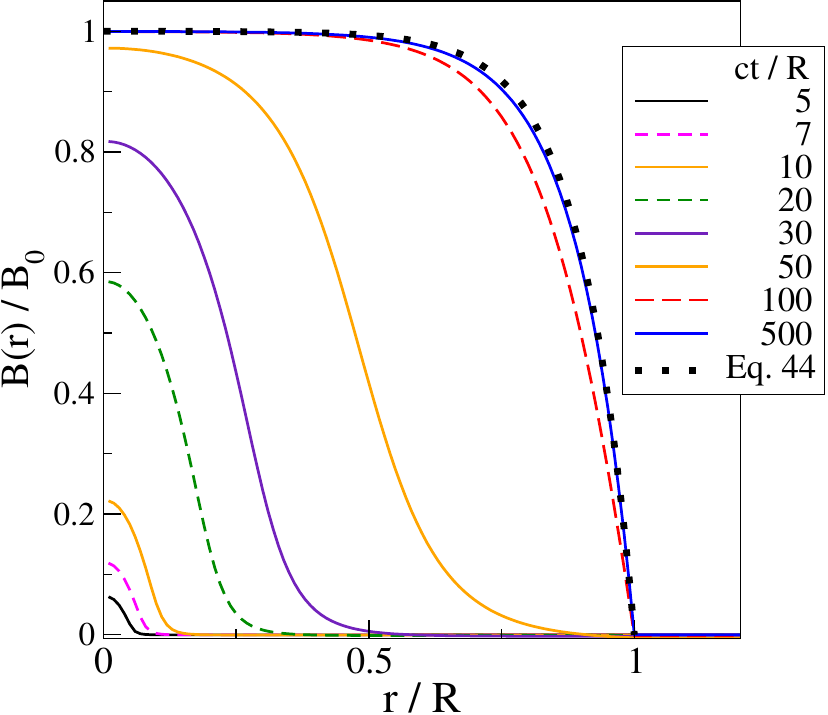}
\caption{Temporal evolution of the magnetic field profile in the
  rotating cylinder (for $\lambda/R=0.1$) for the condensate fraction
  described by the smeared step function
  $f(r,t)=(vt)^\alpha/((vt)^\alpha+r^\alpha)$ with velocity $v=0.01c$
  and exponent $\alpha=10$. }
\label{fig:rot_nucleus}
\end{figure}

\section{Rotate-first protocol with nucleation}
In this Appendix we solve the same problem as in Section~IV.B, with
the only difference that the function $f(r,t)$ which
enters~\eqref{eq:current_rot_prot2} changes. Instead of a
homogeneously growing condensate fraction $f(t)=\tanh(t/\tau)$, here
we assume that superconductivity nucleates in the vicinity of the
cylinder axis, acquires phase coherence there, and afterwards the
superconducting nucleus grows with velocity $v\ll c$ according to
$f(r,t)=\theta(vt-r)$, where $\theta(x)$ is the Heaviside function.

The results are shown in Fig.~\ref{fig:rot_nucleus}. As expected, in
the initial stages the magnetic field is generated essentially only in
the region where the cylinder is superconducting. In the long-time
limit, the magnetic field profile approaches
Eq.~\eqref{eq:B_rotating}, exactly is in the case when the condensate
is switched on homogeneously in the whole cylinder.

\section{Rotating hollow cylinder}
This Appendix is motivated by the experimental study of the
Becker-London effect in a hollow tin cylinder
\cite{Hendricks71}. Since a hollow cylinder is multiply connected, the
phase of the condensate is not uniquely defined and it can exhibit any
integer winding number $n$. Hirsch argues that the conventional theory
does not offer any dynamical explanation for the experimental value of
$n$ in the rotate-first protocol other than energy
minimization~\cite{Hirsch19annalen}. Our goal is to refute this claim.

For the sake of completeness, let us first discuss the cool-first
protocol. In this case the initial state (cylinder at rest and in zero
field) has a well-developed condensate with winding number $n=0$. The
winding number obviously can not change when the cylinder is set into
rotation.  We have checked that, in agreement with the predictions of
Hirsch \cite{Hirsch19annalen}, in such case the magnetic field which
is generated inside the rotating cylinder as well as in its cavity is
$B_0$ (not shown), similarly as in the case of a cylinder without a
hole.

The rotate-first protocol is much more interesting. As a result of the
condensate-formation process, in principle any winding number $n$ can
be generated at the initial stage of this protocol.  Once a condensate
with the winding number $n$ is formed in the sample, then, making use
of~\eqref{eq:London}, the expression for the total current inside the
cylinder~\eqref{eq:current_rot_prot2} should be replaced by
\begin{equation}
\mu_0(j_{\rm ext}+j_s)=\frac{f(t)}{\lambda^2}
\left[\frac{B_0 r}{2} -A-\frac{\hbar}{2e}\frac{n}{r}\right].
\label{eq:current_rot_hollow}
\end{equation}
Note that the current is finite only for $R_1<r<R_2$, and the current
distribution exhibits two jumps: at the inner rim $r=R_1$ of the
hollow cylinder and at the outer rim $r=R_2$.  For definiteness, let
us assume that the superfluid fraction is switched on according to the
formula $f(t)=\tanh(t/\tau)$.

Since the rotate-first protocol starts from an initial state in which
the fields $B$, $E$, and $A$ are zero, one can assume that in the
initial stages all these fields are small. Therefore, in the vicinity
of $r=R_1$ and for $t\ll\tau$, the current distribution can be
approximated by
\begin{equation}
j(r,t)=j_1 \frac{t}{\tau}\theta(t)\theta(r-R_1),
\end{equation}
where $\theta(x)$ is the Heaviside function and 
\begin{equation}
\mu_0 j_1 =\frac{1}{\lambda^2}
\left[\frac{B_0 R_1}{2} -\frac{\hbar}{2e}\frac{n}{R_1}\right].
\label{eq:hollow_current1}
\end{equation}
The magnetic field generated by such a current distribution can be
easily shown to be given by
\begin{equation}
  B(r,t)=\frac{\mu_0 j_1}{4\tau c} \left(ct-|r-R_1|\right)^2.
\label{eq:peak_up}  
\end{equation}
This expression is valid only for $|r-R_1|<ct$, i.e. at short
distances from the inner rim of the cylinder.  In other words, a
narrow peak appears at $r=R_1$ in the magnetic field
distribution. Note that the amplitude of the peak is proportional to
$j_1$ and, therefore, it is controlled by the initial value of $n$.

\begin{figure}[t]
\includegraphics[width = 7. cm]{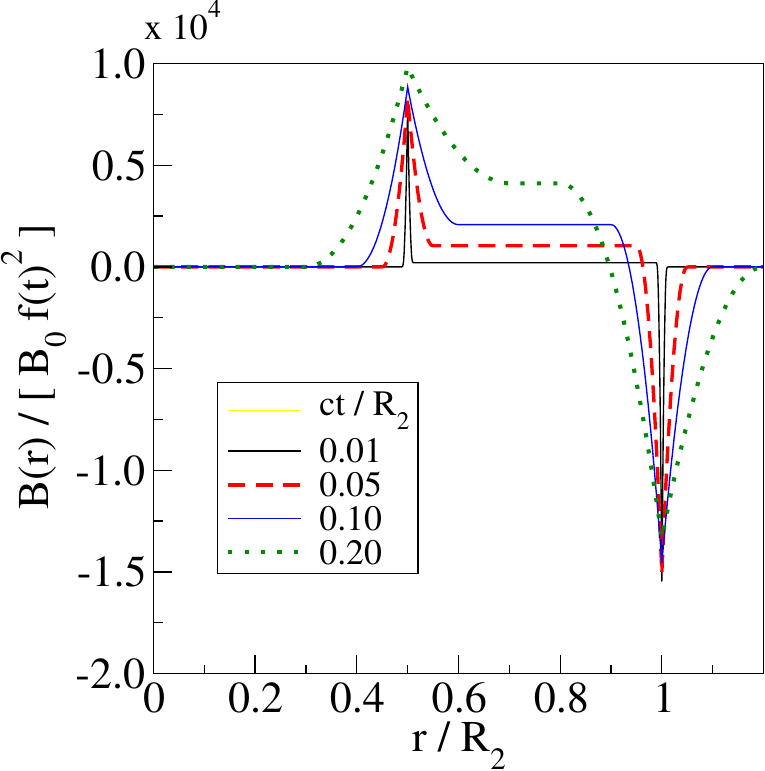}
\caption{Temporal evolution of the magnetic field profile in a hollow
  rotating cylinder with winding number $n=0$ (for $R_1/R_2=0.5$,
  $\lambda/R_2=0.02$, and $c\tau/R_2=50$) in the limit of small times.}
\label{fig:hollow_peaks}
\end{figure}

In the vicinity of $r=R_2$ and for $t\ll\tau$, by the same argument we
find 
\begin{equation}
B(r,t)=-\frac{1}{\lambda^2}
\left[\frac{B_0 R_2}{2} -\frac{\hbar}{2e}\frac{n}{R_2}\right]
\frac{\left(ct-|r-R_2|\right)^2}{4\tau c}.
\label{eq:peak_down}
\end{equation}
As shown in Fig.~\ref{fig:hollow_peaks}, the
estimates~(\ref{eq:peak_up},\ref{eq:peak_down}) agree reasonably well
with the numerically obtained results up to time $t\lesssim t_{\rm
  max}=(R_2-R_1)/(2c)$.

The hollow cylinder studied in \cite{Hendricks71} has inner radius
$R_1=7.1$~mm and outer radius $R_2=10.3$~mm.  Note
that, since the zero-temperature penetration depth of tin is
$\lambda\approx 42$~nm, not too close to the transition temperature we
can assume that $R_2-R_1\gg \lambda$. 
The maximal rotation frequency was
$\omega=2\pi\times 250$~s$^{-1}$, corresponding to a
Becker-London field of $B_0=1.8\times 10^{-8}$~T. This field is many
orders of magnitude smaller than the zero-temperature critical field
of tin, $B_c\approx 3.0\times 10^{-2}$~T. Therefore, as already
mentioned, the transition from the normal to the superconducting state
should be essentially of second order. Assuming that the value of
$f(t)=\tanh(t/\tau)$ is controlled by the changing temperature, we
therefore estimate that the critical field of the cylinder at time $t$
is $B^\ast(t)=B_cf(t)$.

Let us compare now the peak values $B(R_i,t)$ with
$B^\ast(t)$. Assuming for definiteness that the initial winding number
$n=0$, we obtain
\begin{equation}
\frac{|B(R_i,t)|}{B^\ast(t)}=
\frac{B_0 R_i c t}{8\lambda^2 B_c},
\end{equation}
showing that this ratio grows as a function of time. For time
$t=t_{\rm max}$, when the estimate of $B(R_i,t)$ is still
qualitatively correct, we find
\begin{equation}
\frac{|B(R_i,t_{\rm max})|}{B^\ast(t_{\rm max})}\approx
\frac{B_0 R_i (R_2-R_1)}{16B_c \lambda^2}\sim 10^2,
\label{eq:hollow_estimate1}
\end{equation}
where the numerical estimate applies to experimental data from
\cite{Hendricks71}. This means that, if the initial winding number is
$n=0$, the magnetic field becomes larger than the instantaneous
critical field very early in the initial stage of the condensate
formation.

The absolute value of $\mu_0j_1$ (and therefore of the field
$|B(R_1,t)|$ as well) is minimized for a winding number close to
$n_1=\Phi_1/\Phi_0$, where $\Phi_1=\pi R_1^2 B_0$ is the flux
of the Becker-London field through the cavity of the cylinder. For the
experiment~\cite{Hendricks71}, we estimate $n_1\approx 1420$.
Since the winding number is an integer, according
to~\eqref{eq:hollow_current1} the minimal absolute value of $\mu_0j_1$
is of the order of $\Phi_0/(2\pi\lambda^2 R_1)$, where $\Phi_0$ is the
flux quantum.  Making use of this estimate we find that
\begin{equation}
\frac{|B(R_1,t_{\rm max})|}{B^\ast(t_{\rm max})}\sim
\frac{\Phi_0}{16\pi\lambda^2 B_c}\frac{R_2-R_1}{R_1}\sim 0.3.
\end{equation}
This means that $|B(R_1,t_{\rm max})|$ can stay smaller than
$B^\ast(t_{\rm max})$ only if the winding number is $n_1\pm 3$,
i.e. very close to $n_1$. Similar considerations apply also to
$|B(R_2,t_{\rm max})|$, in which case $n_2=\pi R_2^2 B_0/\Phi_0\approx
3000$.

What happens when the magnetic field in the peaks $|B(R_i,t)|$ is
larger than $B^\ast(t)$? In order to answer this question, let us
first note that the magnetic field exhibits a peak at the outer rim
even in a solid cylinder.  If the superconductor is sufficiently
clean, the condensate formation does not seem to be affected by this
fact and the rotating cylinder generates a Becker-London field which
agrees with theory \cite{Hendricks71}. However, this is not the case
in a dirty solid cylinder, where essentially no field is generated by
rotation \cite{Hendricks71}.

Here we take the point of view that both, the dirty solid cylinder and
the hollow cylinder, behave in the same way in the experiment
\cite{Hendricks71}, namely they enter into some sort of the mixed
state.  Below we present some speculations on what happens in both
cases.

As is well known, the phenomenology of the mixed state is quite
complicated, but to some extent it resembles the physics of vortices
in type-II superconductors. For this reason let us have a look at what
would happen in a type-II superconductor with the lower critical field
$B_{c1}(t)=B^\ast(t)$.

{\it Dirty solid cylinder}.  Let us first study the rotating dirty
solid cylinder. In this case, since the superconductor is simply
connected, the winding number of the condensate is initially $n=0$.
According to~\eqref{eq:hollow_estimate1} the field $|B(R_2,t_{\rm
  max})|$ at the rim of the cylinder exceeds $B^\ast(t_{\rm max})$ and
negatively charged vortices (i.e., vortices with magnetic field
pointing along $-z$) are generated here. (More precisely, at the
initial stages of the rotate-first protocol, we should speak about
phase singularities and not about vortices, since the effective
penetration length is huge and the vortex carries less than a flux
quantum in a finite sample.) These vortices are driven by the Lorentz
force in the direction of ${\bf j}\times (-{\bf \hat{z}})$ where ${\bf
  j}$ is given by~\eqref{eq:current_rot_hollow}, i.e. towards the
center of the cylinder. Note that the winding number of the
condensate, when evaluated along a path encircling a negatively
charged vortex, is $+1$. Therefore the winding number evaluated along
a circle with radius $r$ around the center of the cylinder depends on
the number of encircled vortices and is $r$-dependent, $n=n(r)$. The
vortex generation process will end when the total number of vortices
in the cylinder is given by $n(R_2)\approx n_2=\pi R_2^2B_0/\Phi_0$.

At the later stages of the cooling process, when the condensate
amplitude is already well developed, the vortex positions (i.e., the
phase field) remain fixed. Assuming for simplicity that the vortex
distribution is homogeneous, $n(r)=\pi r^2B_0/\Phi_0$, one finds
readily that the total current~\eqref{eq:current_rot_hollow} is
identically zero and no net magnetic flux is generated inside the
rotating cylinder. When the rotation is stopped, a net magnetic flux
$\Phi=-\pi R_2^2B_0$ remains frozen in the sample for sufficiently
strong pinning. Both of these conclusions agree with the experimental
result for a dirty tin cylinder \cite{Hendricks71}. It is worth
pointing out that, surprisingly, our result is quite similar to the
phenomenology suggested by Hirsch~\cite{Hirsch19annalen}.

\begin{figure}[t]
\includegraphics[width = 7. cm]{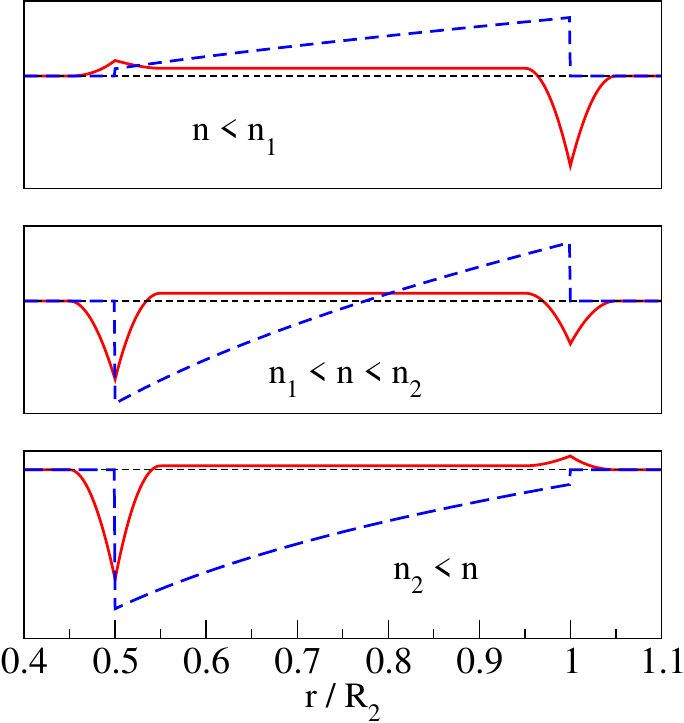}
\caption{Profiles of the magnetic field (full lines) and of the total
  current density~\eqref{eq:current_rot_hollow} (dashed lines) in a
  hollow rotating cylinder for three regimes of the winding number $n$
  in the limit of small times (arbitrary units). The Lorentz force
  acting on the vortices injected at the inner and outer rim is
  proportional to ${\bf j}\times {\bf B}$ and is directed in all cases into
  the sample interior.}
\label{fig:hollow_directions}
\end{figure}

{\it Hollow cylinder}. In what follows, we consider three cases of the
cooling process, depending on the initial value of the winding number
$n$, see Fig.~\ref{fig:hollow_directions}.

(i) If the initial winding number is $n<n_1$, positively charged
vortices enter the sample at the inner rim of the hollow cylinder. By
the Lorentz force they are driven to the outer rim, where they exit
the sample. As a result of this so-called phase-slip process
\cite{Anderson66}, the winding number increases by $+1$. Similarly,
negatively charged vortices enter the sample at the outer rim, and,
being driven by the Lorentz force, they exit the sample at the inner
rim, again increasing the winding number by $+1$.

Both of these processes continue until $n\approx n_1$. At that point,
the peak of the magnetic field at $r=R_1$ disappears, but negatively
charged vortices are still created at the outer rim, since $n_2>n_1$.
Similarly as in the case of a solid cylinder, these vortices remain
trapped inside the superconductor and the resulting vortex matter can
be described by an $r$-dependent winding number $n_v(r)=\pi
(r^2-R_1^2)B_0/\Phi_0$.  Therefore the total $r$-dependent winding
number is $n(r)=n_1+n_v(r)=\pi r^2B_0/\Phi_0$. Note that $n(R_2)=n_2$,
as is in fact required to stop the vortices from entering at the outer
rim.

(ii) If the initial winding number is $n>n_2$, the signs of the
currents and magnetic fields are reversed with respect to (i). As a
result, positively charged vortices move from the outer to the inner
rim and negatively charged vortices move in the opposite direction.
This means that the phase slip processes decrease the winding number
by $-1$. When the winding number reaches $n=n_2$, the positively
charged vortices do not enter the sample any more at the outer
rim. However, the negatively charged vortices do continue entering the
sample from the inner rim, generating vortex matter inside the
superconductor. One checks readily that the resulting vortex matter
can be again described by the winding number $n_v(r)=\pi
(r^2-R_1^2)B_0/\Phi_0$, thereby reducing the winding number at the
inner rim to $n_2-n_v(R_2)=n_1$, in agreement with the results found
in case (i).

(iii) If the initial winding number is $n_1<n<n_2$, negatively charged
vortices are generated at both rims of the cylinder, which are driven
into the interior of the sample by the Lorentz force. At the outer
rim, these processes lead to an increase of the $r$-dependent winding
number, until finally $n(R_2)=n_2$. Similarly, at the inner rim the
$r$-dependent winding decreases, until $n(R_1)=n_1$. The resulting
profile of $n(r)$ is again the same as in case (i).

What does this result imply? Since $n(r)=\pi r^2B_0/\Phi_0$, the total
current~\eqref{eq:current_rot_hollow} is identically zero and the
magnetic flux is generated neither in the superconductor, nor in its
cavity. However, when the rotation is stopped, a net magnetic flux
$\Phi_{\rm sample}=-\pi (R_2^2-R_1^2)B_0$ remains frozen in the sample
for sufficiently strong pinning. Moreover, the winding number $n_1$
generates a magnetic flux $\Phi_{\rm cavity}=-n_1\Phi_0=-\pi R_1^2
B_0$ in the cavity. These results are again consistent with the
findings of~\cite{Hendricks71}.

{\it Conclusion}. We have demonstrated that, provided that the dirty
and hollow cylinders behave similarly to type-II superconductors, the
London electrodynamics is able to describe the experimental results
in~\cite{Hendricks71}. In particular, we were able to identify the
mechanism by which a hollow cylinder chooses the correct winding
number $n$.


\begin{thebibliography}{99}
\bibitem{Hirsch20}J. Hirsch, {\it Superconductivity begins with H},
  World Scientific, New Jersey, 2020.
  
\bibitem{Schrieffer64}J.R. Schrieffer, {\it Theory of
  Superconductivity}, W.A. Benjamin, New York, 1964.

\bibitem{deGennes66}P.G. de Gennes, {\it Superconductivity in Metals
  and Alloys}, W.A. Benjamin, New York, 1966.
  
\bibitem{Tinkham04}M. Tinkham, {\it Introduction to
  Superconductivity}, 2$^{\rm nd}$ Ed., Dover, New York, 2004.

\bibitem{BCS11}{\it BCS: 50 Years}, ed. by L.N. Cooper and D. Feldman,
  World Scientific, New Jersey, 2011.
  
\bibitem{Hirsch16annals}J.E. Hirsch, Annals of Physics {\bf 373}, 230
  (2016).

\bibitem{Hirsch20epl}J.E. Hirsch, EPL {\bf 130}, 17006 (2020).

\bibitem{Hirsch25}J.E. Hirsch, Physica C
%Physica C: Superconductivity and its applications 
{\bf 629}, 1354618 (2025).

\bibitem{Hirsch14scripta}J.E. Hirsch, Phys. Scr. {\bf 89}, 015806
  (2014).

\bibitem{Hirsch19annalen}J.E. Hirsch, Ann. Phys. (Berlin) {\bf 531},
  1900212 (2019).

\bibitem{Meissner33}W. Meissner and R. Ochsenfeld, Naturwissenschaften
  {\bf 21}, 787 (1933).

\bibitem{Tate89}J. Tate, B. Cabrera, S.B. Felch, and J.T. Anderson,
  Phys. Rev. Lett. {\bf 62}, 845 (1989).

\bibitem{Verheijen90}A.A. Verheijen, J.M. van Ruitenbeek, R. de Bruyn
  Ouboter, L.J. de Jongh, Nature (London) {\bf 354}, 418 (1990).

\bibitem{Sanzari96}M.A. Sanzari, H.L. Cui, F. Karwacki,
  Appl. Phys. Lett. {\bf 68}, 3802 (1996).

\bibitem{Becker33}R. Becker, G. Heller, and F. Sauter, Z. Phys. {\bf
  85}, 772 (1933).

\bibitem{London61}F. London, {\it Superfluids}, Vol. I, Dover, New
  York, 1961.

\bibitem{Nikulov22}A. Nikulov, 
%The law of entropy increase and the Meissner effect, 
Entropy {\bf 24}, 83 (2022).

\bibitem{Koizumi20}H. Koizumi, EPL {\bf 131}, 37001 (2020).

\bibitem{Koizumi21}H. Koizumi, J. Supercond. Nov. Magn. {\bf 34}, 1361
  (2021).

\bibitem{London48}F. London, Phys. Rev. {\bf 74}, 562 (1948).

\bibitem{Ginzburg50}V.L. Ginzburg and L.D. Landau, Zh. Eksp.
  Teor. Fiz. {\bf 20}, 1064 (1950).

\bibitem{note_GL}We would like to point out that, although this does
  not appear to be widely appreciated, the term involving the gradient
  of the $\theta({\bf x},t)$ field was introduced already at least in
  \cite{London48}. The later work by Ginzburg and Landau
  \cite{Ginzburg50} provided further justification
  of~Eq.~\eqref{eq:London}.
  
  \bibitem{Hirsch03}J.E. Hirsch, Phys. Rev. B 68, 184502 (2003).

\bibitem{Jackson99}J.D. Jackson, {\it Classical Electrodynamics},
  3$^{\rm rd}$ Ed., J. Wiley, New York, 1999.

\bibitem{Hendricks71}J.B. Hendricks, C.A. King, and H.E.Rorschach,
  J. Low Temp. Physics {\bf 4}, 209 (1971).  

\bibitem{Kozhevnikov21}V. Kozhevnikov, J. Supercond. Nov. Magn. {\bf
  34}, 1979 (2021).

\bibitem{note_conserv_en}When evaluating $j_sE\propto f\partial
  A^2/\partial t$, we have made use of the identity $f\partial
  A^2/\partial t=\partial (fA^2)/\partial t-A^2\partial f/\partial
  t$. The first term contributes to $\Delta U$, while the second one
  to $W_B$.

\bibitem{note_W_B}This result can be checked by an explicit evaluation
  of~\eqref{eq:W_B}, assuming that for large $\tau$ the magnetic field
  is given at all times by~\eqref{eq:meissner_plate}, where $\lambda$
  is replaced by $\lambda_{\rm eff}(t)$.

\bibitem{Livingston69}J.D. Livingston and W. DeSorbo, in {\it
  Superconductivity}, Vol.~2, ed. by R.D. Parks, Marcel Dekker, New
  York, 1969.

\bibitem{Hirsch_heat}Our estimate of $Q_J$ is essentially the same as
  the one obtained by Hirsch in \cite{Hirsch16annals}. In that paper,
  he compares this result with experimental data for the Meissner
  effect in an ellipsoidal sample of Sn [W.H. Keesom and P.H. van
    Laer, Physica {\bf 4}, 487 (1937)]. From the observed small value
  of the ratio $Q_J/(T\Delta S)$ he concludes that $B_0=(1+p)B_c$ with
  $p<3.3\times 10^{-4}$.  He argues that this estimate is incompatible
  with the observed duration $t_{\rm max}$ of the process, since the
  theoretical estimate $t_{\rm max}\approx \mu_0\sigma L^2/(2p)$ is
  too large.  We believe that this conclusion is not valid, since in
  samples with a small demagnetizing factor ${\cal D}$, one should use
  the estimate $t_{\rm max}\approx \mu_0\sigma L^2/(2p+{\cal D})$
  \cite{Livingston69}.  Taking the appropriate value ${\cal D}\approx
  0.088$, we find that the estimated time $t_{\rm max}$ is 130 times
  smaller than Hirsch's estimate.

\bibitem{Hirsch_entropy}Another argument raised by Hirsch
  \cite{Hirsch20epl} is as follows. When the temperature of a
  superconducting plate at temperature $T$ in contact with a reservoir
  at temperature $T_0<T$ decreases by $\Delta T$, the Joule heat
  generated in the transition process is proportional to $\Delta T$,
  whereas thermodynamics is consistent with an entropy increase only
  proportional to $(\Delta T)^2$. This argument is incorrect for two
  reasons: first, the Joule heat is not extensive, being produced only
  in a surface layer of width $\lambda$. Second, a homogeneous
  temperature profile inside the superconductor is assumed, which,
  according to III.B, leads to an overestimate of the Joule heat.

\bibitem{Darwin36}C.G. Darwin, Proc. Roy. Soc. A {\bf 154}, 61 (1936).

\bibitem{Geurst93}J.A. Geurst, H. van Beelen, Physica C {\bf 208}, 43
  (1993).

\bibitem{Lipavsky13}P. Lipavsk\'{y}, J. Bok, and J. Kol\'{a}\v{c}ek, Physica
  C {\bf 492}, 144 (2013).

\bibitem{Reppy65}J.D. Reppy, Phys. Rev. Lett. 14, 733 (1965).

\bibitem{Goodstein85}D.L. Goodstein, {\it States of Matter}, Dover,
  New York, 1985.

\bibitem{Aharonov59}Y. Aharonov and D. Bohm, Phys. Rev. {\bf 115}, 485
  (1959).

\bibitem{note_KG}Strictly speaking, this is true only in situations
  when the condensate fraction $f$ does not depend on the spatial
  coordinate.

\bibitem{Kibble76}T.W. Kibble, J. Phys. A: Math. Gen. {\bf 9}, 1387
  (1976).

\bibitem{Zurek85}W.H. Zurek, Nature {\bf 317}, 505 (1985).

\bibitem{Lee24}K. Lee, S. Kim, T. Kim, and Y. Shin, Nature Physics
  {\bf 20}, 1570 (2024).

\bibitem{Anderson66}P.W. Anderson, Rev. Mod. Phys. {\bf 38}, 298
  (1966).

\end{thebibliography}
\end{document}